\documentclass[preprint,12pt,numbers,sort&compress]{elsarticle}

\usepackage{amssymb}
\usepackage{amsthm}
\usepackage{color}

\usepackage{float}
\graphicspath{{fig/}}
\usepackage{multirow}
\usepackage{subfigure}
\usepackage{graphicx}
\usepackage{mathrsfs}

\journal{Experimental Thermal and Fluid Science}

\begin{document}

\begin{frontmatter}



\title{Inlet temperature driven supercritical bifurcation of combustion instabilities in a Lean Premixed Prevaporized combustor}


\author[buaa,imperial]{Xiao~Han} 
\author[imperial]{Davide~Laera} 
\author[imperial]{Aimee S. Morgans} 
\author[buaa]{Yuzhen~Lin} 
\author[buaa]{Chi~Zhang\corref{cor1}} 
\author[buaa]{Xin~Hui}
\author[uc]{Chih-Jen~Sung} 

\address[buaa]{National Key Laboratory of Science and Technology on Aero-Engine Aero-thermodynamics, Co-innovation Center for Advanced Aero-Engine, School of Energy and Power Engineering, Beihang University, Beijing, 100083, P. R. China\\}

\address[imperial]{Department of Mechanical Engineering, Imperial College London, London SW7 2AZ, UK}

\address[uc]{Department of Mechanical Engineering, University of Connecticut, Storrs, CT 06269, USA} 

\cortext[cor1]{Corresponding author: zhangchi@buaa.edu.cn}

\begin{abstract}

The present article reports experimental observation and analyses of a supercritical bifurcation of combustion instabilities triggered by the air inlet temperature ($T_a$). The studies are performed with a pressurised kerosene fuelled Lean Premixed Prevaporized (LPP) combustor operated under elevated temperature. Unlike some previous studies, starting from an unstable condition of the system, the amplitude of combustion instabilities suddenly decrease when $T_a$ exceeds a critical value of $T_a$=570 K. When the temperature is lowered back the system returns to being unstable without featuring any hysteresis behaviour, as expected in case of a supercritical bifurcation. The unstable flames feature a periodic axial motion of lift-off and re-ignition, characterized as Helmholtz mode. The phase difference between chemiluminescence and pressure signals is found to increase with $T_a$, exceeding 90 degrees (out of phase) for temperatures higher than 570 K. A low order network framework is conducted, illustrating that when $T_a$ is increased a stability shift of this mode is predicted at $T_a$ near 570 K, in good agreement with the experimental observations. The impact of $T_a$ on the spray characteristics is also examined, finding that higher $T_a$ promotes fuel evaporation and reduces equivalence ratio fluctuation at the exit of the swirler.

\end{abstract}


\end{frontmatter}

%\linenumbers

%\section{}
%\label{}





\section{Introduction}

%%%%%%%%%%%%%%%%%%%%%%%%%%%%%%%%%%%%%%%%%%%%%%%%%%%%%%

%%%%%%%%%%%%%%%%%%%%%
%                                 %
%.          INTRODUCTION                 %
%                                 %
%%%%%%%%%%%%%%%%%%%%%

Modern gas turbines for power generation and aero-engines are usually operated with lean premixed prevaporized (LPP) combustors to reduce the NOx emissions~\cite{li2016emission}. One of the main design challenges of the lean combustion process is to suppress instabilities that arise in this type of device, these typically result in a resonant coupling between pressure fluctuations ($p'$) and heat release rate oscillations ($\dot q'$) produced by the combustion~\cite{rayleigh1878explanation}. This kind of coupling is unfortunately promoted in these configurations as they feature flame properties that are more sensitive to perturbations. Combustion instabilities are an undesirable behaviour that needs to be avoided since they have many detrimental effects such as augmented vibrations, increased heat load to the combustor walls, flame flashback or blowout and they lead in extreme cases to mechanical failure~\cite{lieuwen2005combustion, al2015review}. 

\begin{figure}[htbp]
\centering
\includegraphics[width=0.7\textwidth]{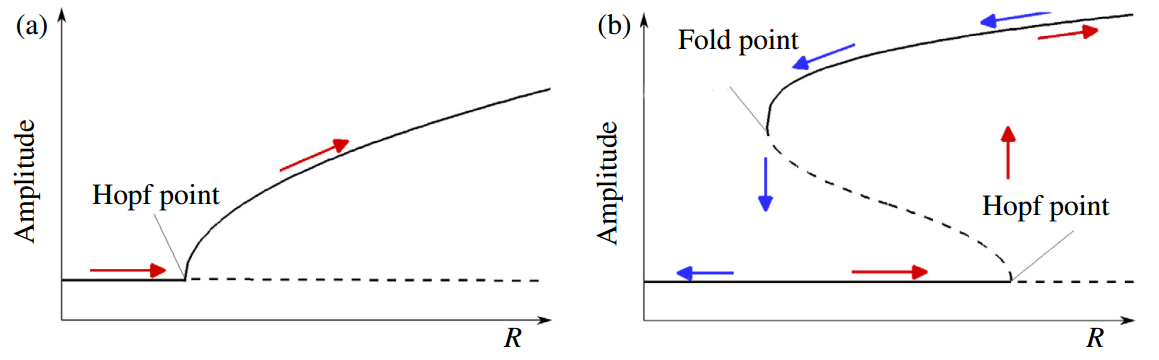}
\caption{Examples of bifurcation diagrams with the variation of a control parameter R. (a) Supercritical bifurcation and (b) subcritical bifurcation \cite{laera2017finite}. As the control parameter R is increased, the system follows the red arrow path. As it is decreased, the system follows the blue arrow path. (For interpretation of the references to colour in this figure legend, the reader is referred to the web version of this article.)}
\label{bifurcation} 
\end{figure}

In practice, combustion instabilities of a given combustor are highly sensitive to many operating conditions, such as equivalence ratio~\cite{ebi2017flame}, stratification ratio~\cite{han2019flame}, inlet mass flow rate~\cite{yoon2013effect, kim2019experimental}, temperature~\cite{broda1998experimental, huang2004bifurcation}, etc. It has been previously observed that oscillations amplitudes can change dramatically when combustor operation parameters pass through some critical values~\cite{huang2009dynamics}. In dynamical system theory, this behaviour is usually called "bifurcation" and the parameter values at which they occur is called the bifurcation point (or Hopf point)~\cite{strogatz2018nonlinear}. Figure~\ref{bifurcation} shows two diagrams, describing the bifurcation dynamics as a function of a control parameter R. The nonlinear behaviour around the Hopf bifurcation point determines two different types of bifurcation: Fig.~\ref{bifurcation}(a) shows a supercritical bifurcation, which is characterised by a gradual increase of the amplitude when the control parameter reaches the bifurcation point. The second bifurcation type shown in  Fig.~\ref{bifurcation}(b) is a subcritical bifurcation which features a sudden jump when the control parameter exceeds the Hopf bifurcation value. Upon reaching the limit cycle equilibrium, the system stays in the high amplitude branch even for values of R lower than the one corresponding to the Hopf point, until the fold point is reached.

The above-described bifurcation phenomena for a thermoacoustic system have been studied numerically and experimentally in recent decades with a focus on gaseous laminar flames~\cite{Kashinath2014Nonlinear}, swirling turbulent flames~\cite{broda1998experimental,huang2004bifurcation} in longitudinal systems, such as a Rijke tube~\cite{kabiraj2012bifurcations,weng2016investigation}, and recently, also for annular combustors~\cite{prieur2017hysteresis}. According to these studies, a variation of certain operating conditions induces a modification of the nonlinear flame response to the incoming acoustic perturbations, the Flame Describing Function (FDF), the interaction between $p'$ and $\dot q'$ is consequently altered, leading to a drift of the thermoacoustic status of the combustion system. Two physical mechanisms can lead to bifurcations: (1) the change of flame dynamics leading to a modification of the gain of the FDF~\cite{huang2004bifurcation,Moeck2008Subcritical,laera2017finite}; (2) change of convection time, i.e., variation in the FDF phase~\cite{ebi2017flame,janus1996model,Noiray2008A}. These two effects are now discussed below. 

The link between flame shape transitions and the thermoacoustic bifurcation behaviour of a combustor is discussed by multiple authors, e.g., see the experimental work of Broda et al.~\cite{broda1998experimental} and the LES study by Huang et al.~\cite{huang2004bifurcation}. They have found that the instability triggered by increasing the inlet air temperature and the equivalence ratio is linked to a modification of the flame shape which is characterised by a more enhanced penetration of the heat release zone in the corner recirculation zone. Different flame shapes and locations in stable and unstable cases were also observed by Moeck et al.~\cite{Moeck2008Subcritical}. A nonlinear model based on perturbation of equivalence ratio and flame speed was proposed to reproduce the phenomenon. For the impact of the convection time, the first study was performed by Janus et al.~\cite{janus1996model}.  A theoretical model was defined and successfully reproduced the experimental results revealing the importance of the convection time to described inlet temperature-triggered bifurcation phenomena. A recent study of the equivalence ratio-triggered bifurcation also emphasises the importance of convection time, while keeping nearly constant the gain of the FDF in varying conditions~\cite{ebi2017flame}. An instability region prediction method based on the ratio of convection time to the oscillation period $Cn=\tau/T$ has been developed and validated~\cite{lieuwen2001mechanism}. 

%%%%%%%

A smaller group of studies deal with systems where the fuel is injected as a spray, a configuration more common in practical combustors but characterised by an increased level of complexity.  Indeed, the spray dynamics and its evaporation is found to have a direct impact on the flame properties~\cite{giuliani2002influence, de2009investigations,tachibana2015experimental, sidey2018stabilisation}. Some early experiments of hysteresis phenomena in combustion are reported by Providakis et al \cite{providakis2012characterization} showing two different time-averaged flame shapes at the same fuel split ratio when operated with different routines. These two flames have found to be characterised by different thermoacoustic properties~\cite{renaud2015flame} and the transition between the two states is observed to be triggered by acoustic field perturbations~\cite{renaud2017bistable}. As discussed for gaseous flames, large-scale flame dynamics characterise the thermoacoustic oscillations. Both global motions involving flame lift-off and re-ignition~\cite{temme2014combustion} and local flame surface dynamics within the combustor corner region~\cite{dhanuka2011lean} have been observed. Given the intrinsic partially premixed nature of spray flames, the impact of equivalence ratio fluctuations on the thermoacoustic stability is also non-negligible. Han et al.~\cite{han2017combustion} have found with the increase of inlet temperature, the amplitude of combustion instabilities drop non-linearly. 
%This is possibly due to a better mixing is achieved with higher inlet temperature, resulting in a different flame transfer function and a shift of oscillation frequency.
However, due to limited conditions, a comprehensive conclusion was not obtained and the relation between inlet air temperature and combustion instabilities in spray flames still remains unanswered.

In the present study, we aim to fill this gap by presenting experiments and modelling analysis of a kerosene-fuelled LPP combustor. The combustor is operated at elevated pressure and varying inlet temperatures. This article is organised as follows. At first, the experiment setup and results are presented. Spectra of the multiple signals and images of flame dynamics are shown. The phase relationship between the heat release rate and pressure is also analysed. Further analysis is conducted using low order network analysis and spray property estimation.

%%%%%%%%%%%%%%%%%%%%%
%                                 %
%.          EXPERIMENTAL SETUP         %
%                                 %
%%%%%%%%%%%%%%%%%%%%%

\section{Experimental setup}
%%%%%%%%%%%%%%%%%%%%%%%%%%%%%%%%%%%%%%%%%%%%%%%%%%%%%%%%%%%%%
\begin{figure}[htbp]
\centering
\includegraphics[width=1\textwidth]{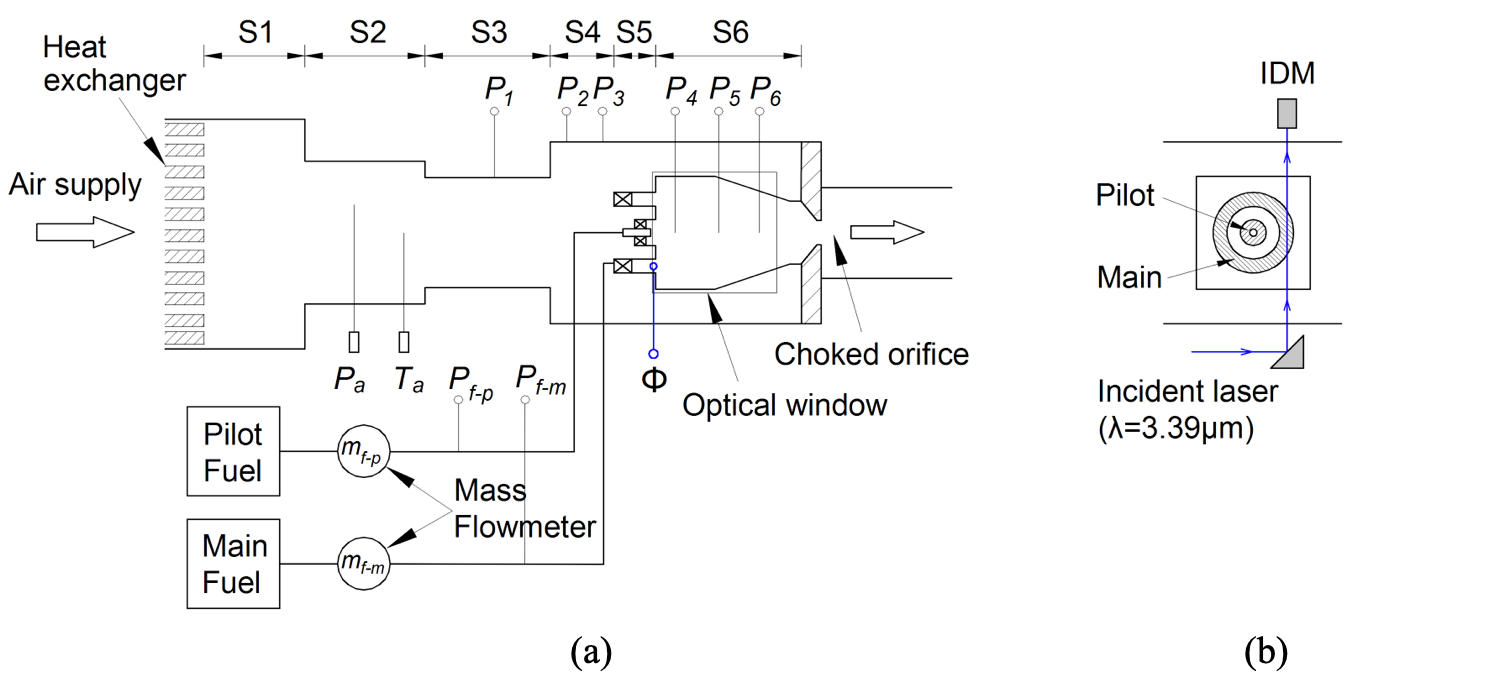}
\caption{Schematic view of the a) combustion test rig with pressure and optical measurements locations indicated, b) the setup of equivalence ratio fluctuation measurements.}
\label{schematic} 
\end{figure}
%%%%%%%%%%%%%%%%%%%%%%%%%%%%%%%%%%%%%%%%%%%%%%%%%%%%%%%%%%%%%
In the present work, a centrally-staged LPP combustor with the pilot stage and main stage is used, fuelled with Chinese kerosene RP-3\footnote{This fuel has properties similar to the Jet-A~\cite{mao2019experimental}.}. Air is injected through two different coaxial swirlers (diagonal and axial) with the main air mass flow passing through the diagonal passage. Further information on this system can be found in Ref~\cite{han2016effect}. 

Experiments are conducted at the High-pressure High-temperature Combustion Test Rig of Beihang University. This rig shown in Fig.~\ref{schematic} is designed for emission and combustion instability experiments of gas turbine combustors operated with a maximum air mass flow rate of 2 kg/s, which could be pre-heated to a maximum temperature of 850 K at 3MPa of total pressure. The air is compressed and stored in high-pressure tanks and then preheated using a heat exchanger without vitiation. The inlet consists of a perforated plate with an aperture ratio of 6\%. The burnt gas exit is choked by an orifice. The inlet air mass flow rate $\dot{m_a}$ is measured by a standard ASME orifice with an accuracy of 1\%, while the fuel for the pilot and main stages of the combustor is measured by two Coriolis force flowmeters with an accuracy of 1\%. The inlet temperature $T_a$ and total pressure are continuously monitored by a thermocouple and a total pressure probe just upstream of the combustor.

Dynamic pressure sensors (PCB 112A22) are installed in positions upstream and within the combustor, as shown in Fig.~\ref{schematic}(a). Two sensors (PCB S112A22) are installed in the fuel supply lines for the pilot and main stage allowing to monitor the pressure fluctuations upstream of fuel injectors. PCB S112A22 is similar to 112A22, but is able to make direct contact with the fuel. The measuring range of the sensors is 0-345 kPa. A photomultiplier (PMT) of Hamamatsu H9306 is used to measure the global CH$ ^* $ chemiluminescence emissions as representative of heat release rate~\cite{hardalupas2004local}. For most of the hydrocarbon molecules, absorption of 3.39 $\mu$m He-Ne laser occurs and can be used as a quantitative method to measure the mole fraction of either liquid or gaseous fuel~\cite{tsuboi1985light}. Spray equivalence ratio fluctuations are measured at the outlet of the main stage (before combustion), as shown within Fig.~\ref{schematic}(a). The incident laser is reflected to pass through the edge of the main stage channel and then received by an infrared detector module (IDM, Hamamatsu C12495-211S),  as shown in Fig.~\ref{schematic}(b). As no calibration is conducted, we only take the data for qualitative analysis in the present paper. The fuel mass flow rates for the two stages obtained by the Coriolis force flowmeters are also sampled. All data are collected by a DAQ system (National Instruments, NI9215) with a sampling frequency $f_s=$8192 Hz and with a total length of 30000 points for each condition. An intensified high-speed camera (Photron, Fastcam SA4) is used to capture the flame dynamics. The image dimension is 120 mm $\times$ 100 mm, corresponding to 768 $\times$ 640 pixels (length $\times$ height). A total of 4659 snapshots are obtained at a sampling rate of 6000 fps. Both the PMT and high-speed camera are equipped with CH$ ^* $ filters (430$ \pm $5 nm).

\begin{table}[ht]
\centering
\caption{Operating conditions for experiments.}
\label{table_op}
\begin{tabular}{ccccc}
\hline
$\dot{m_a}$   & $T_a$           & $P_a$          & \multirow{2}{*}{FAR} & \multirow{2}{*}{SR} \\ \cline{1-3}
kg/s & K           & MPa        &                      &                     \\ \hline
0.5  & 560$\rightarrow$ 617$\rightarrow$ 530 & $\sim$0.50 & 0.021                & 30\%                \\ \hline
\end{tabular}
\end{table}

This work focuses on the effect of inlet temperature on combustion instabilities. During the experiments, the air mass flow rate $\dot{m_a}$ is fixed at 0.5 kg/s, while the inlet temperature $T_a$ varies from 530 K to 617 K, with the operating pressure $P_a$ at about 0.5 MPa. Considering the complexity of the heat exchanger and the large mass flow rate, it is very difficult to increase the inlet temperature continuously. Therefore a step of about 10 K is chosen, which is the minimum step that was achievable. The operating conditions are summarised in Tab.~\ref{table_op}. The fuel to air ratio (FAR) is fixed at 0.021, corresponding to a total equivalence ratio of 0.31 with a staging ratio of 30\%, i.e.,  30\% of the fuel enters through the pilot and the rest is injected in the main stage. The temperature is at first increased from 560 K to 617 K and then decreased back to 530 K to verify the presence of hysteresis phenomena. In total, ten operating conditions with different inlet temperatures have been recorded.

%%%%%%%%%%%%%%%%%%%%%
%                                 %
%.          EXPERIMENTAL RESULTS     %
%                                 %
%%%%%%%%%%%%%%%%%%%%%

\section{Experimental results}
\subsection{Characteristics of combustion oscillations}

\begin{figure}[htbp]
\centering
\includegraphics[width=0.5\textwidth]{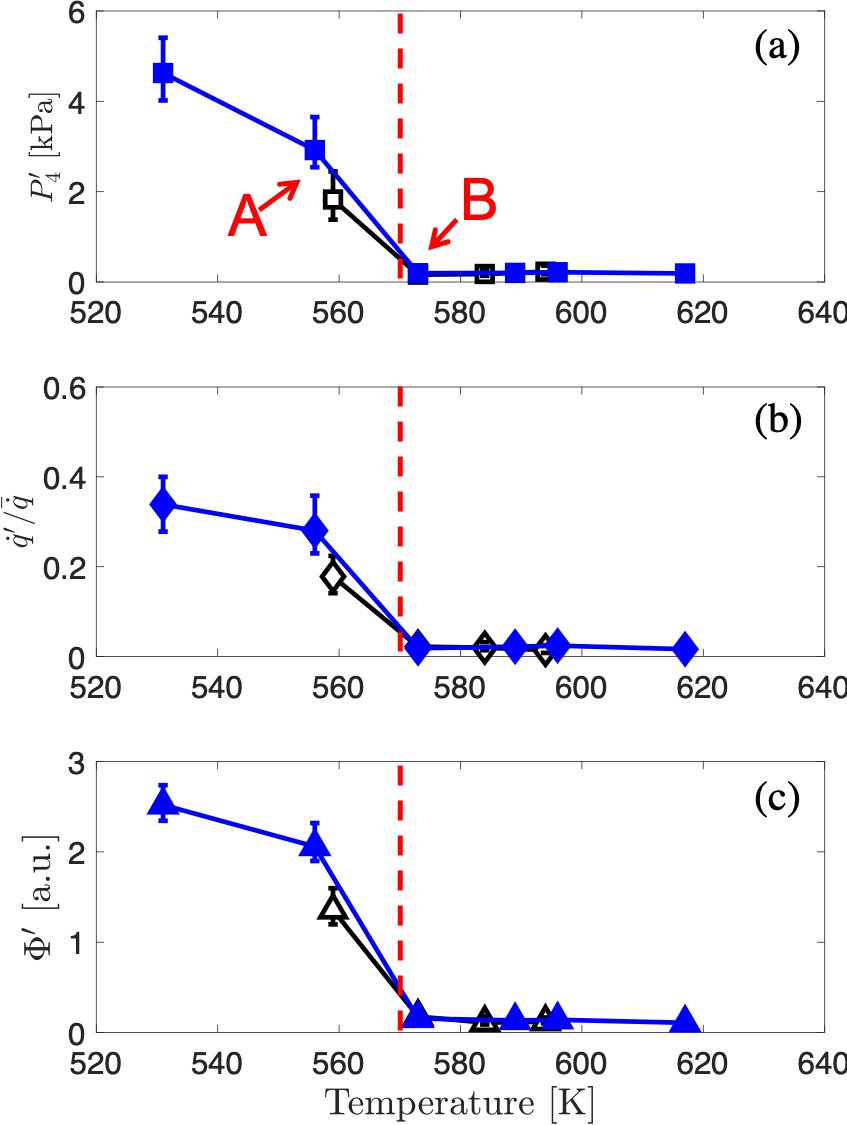} %Bifurcation50_errorbar
\caption{Combustion oscillation amplitude as a function of inlet temperature. (a) Amplitude of the fluctuation of $P_4'$, (b) global intensity of CH* (normalised ) and (c) equivalence ratio (arbitrary scale). The dash line marks the critical inlet temperature of the bifurcation. The hollow black symbols mark the path of increasing inlet temperature, while the solid blue symbols mark the path of decreasing temperature. $A$ and $B$ mark $T_a$=556 K and $T_a$=573 K.}
\label{50}
\end{figure}
 
The system bifurcation map as a function of the air inlet temperature $T_a$ is shown in Fig.~\ref{50}, reporting (a) amplitude of the pressure fluctuations in all tested points, (b) the normalised fluctuation of global intensity of CH* and (c) equivalence ratio. The original data comes from P4 sensor, which is located in the combustor chamber and close to the flame. Chemiluminescence of CH* sampled by PMT is used as the indicator of heat release rate $\dot q'$, while the infrared absorption signal is used as the representative of equivalence ratio $\phi'$. The above data are post-processed via the Fast Fourier transform (FFT) method. Each data is divided into 6 segments and calculated separately. Figure ~\ref{50} is then plotted with the mean amplitude of each parameter. The uncertainties of oscillation are marked as error bars. The scattering of the amplitude is within 35\%, which is normal and also observed in other configurations, as combustion instabilities are nonlinear phenomenon~\cite{han2019flame,kabiraj2012bifurcations,weng2016investigation}. For conditions above 570 K, the error bars are not visible as the amplitude is too small. By varying the inlet air temperature a supercritical bifurcation is clearly observed in Figs~\ref{50}. Inlet air temperature $T_a$ is set at 560 K at first and increases step-by-step to 617 K (marked as the hollow black symbols in Fig.~\ref{50}). Then the $T_a$ drops back to 530 K (marked as the solid blue symbols in Fig.~\ref{50}). It is found with $T_a$=560 K,  the combustor is unstable with a high amplitude of ~1.8 kPa. Once increasing the $T_a$ beyond 570 K, the amplitude drops to near-zero until 617 K. As the $T_a$ decreases back, the amplitude remains low until 570 K is reached. Further reduction of $T_a$ to 530 K leads to a progressive increase of the amplitude of the oscillations, following the similar path back to the previous status, without the hysteresis phenomena (as indicated by the blue arrows in Fig.~\ref{50} plots). This behaviour of combustion instabilities as a function of inlet air temperature meets the description of supercritical bifurcation mentioned above. A Hopf bifurcation point of $T_{air,Hopf}\approx$ 570 K is found (indicated with the dashed vertical line in Fig.~\ref{50}), above which the system stabilises. Finally, strong combustion instabilities characterised by high amplitudes of pressure fluctuations $p'$ together with large oscillations of  $\dot q'$ and $\phi'$ are found when the inlet air temperature is set to the lowest $T_a$ of 530 K. Therefore, two representative test conditions crossing the bifurcation line are chosen for further discussion, as marked with $A$ and $B$ in Fig.~\ref{50}. Condition $A$ is unstable with $T_a$=560 K while $B$ is stable with $T_a$=573 K.

\begin{figure}[htbp]
\centering
     \subfigure[\label{473}]{%
       \includegraphics[width=0.6\textwidth]{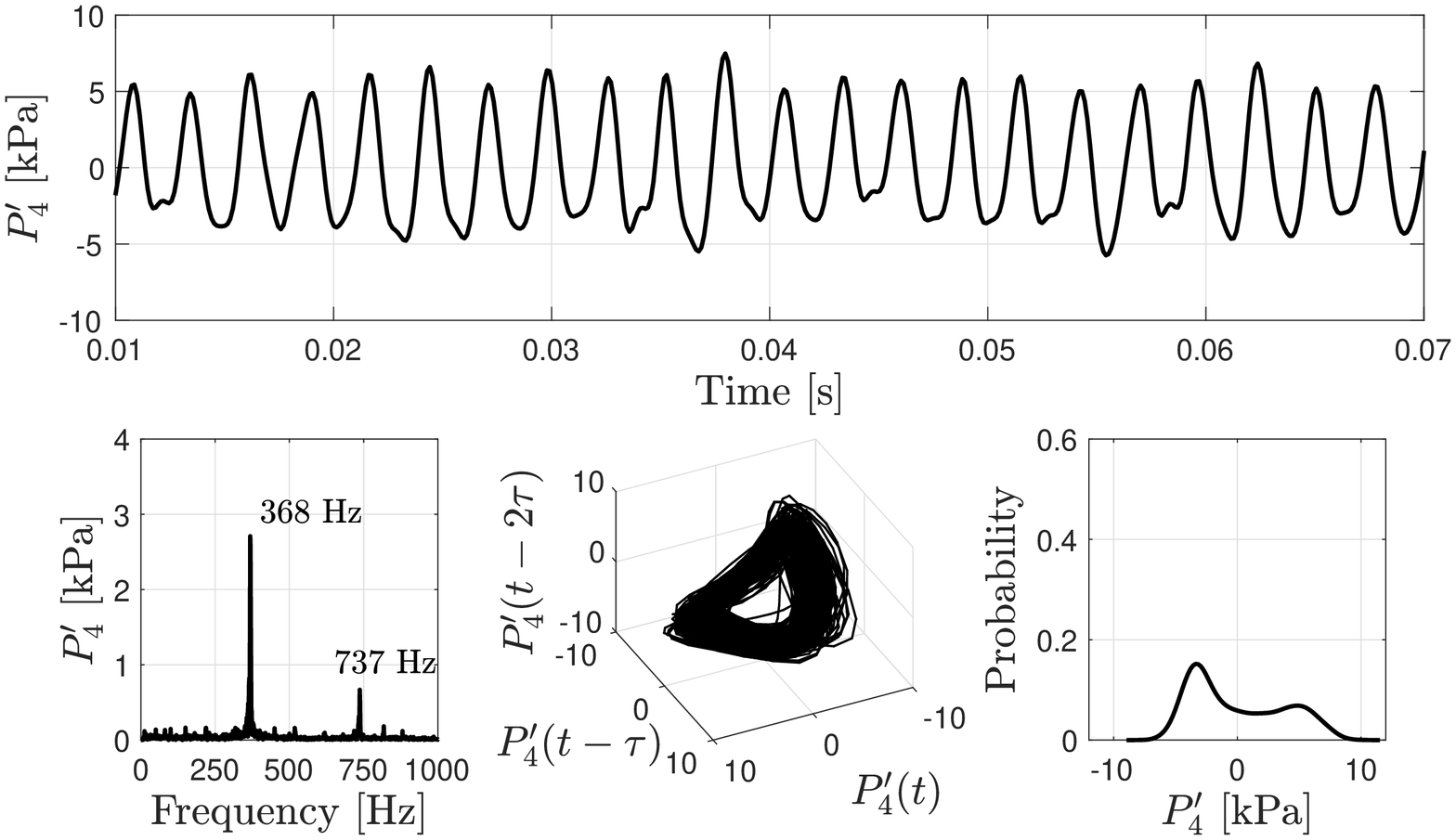}
     }\\
     \subfigure[\label{474}]{%
       \includegraphics[width=0.6\textwidth]{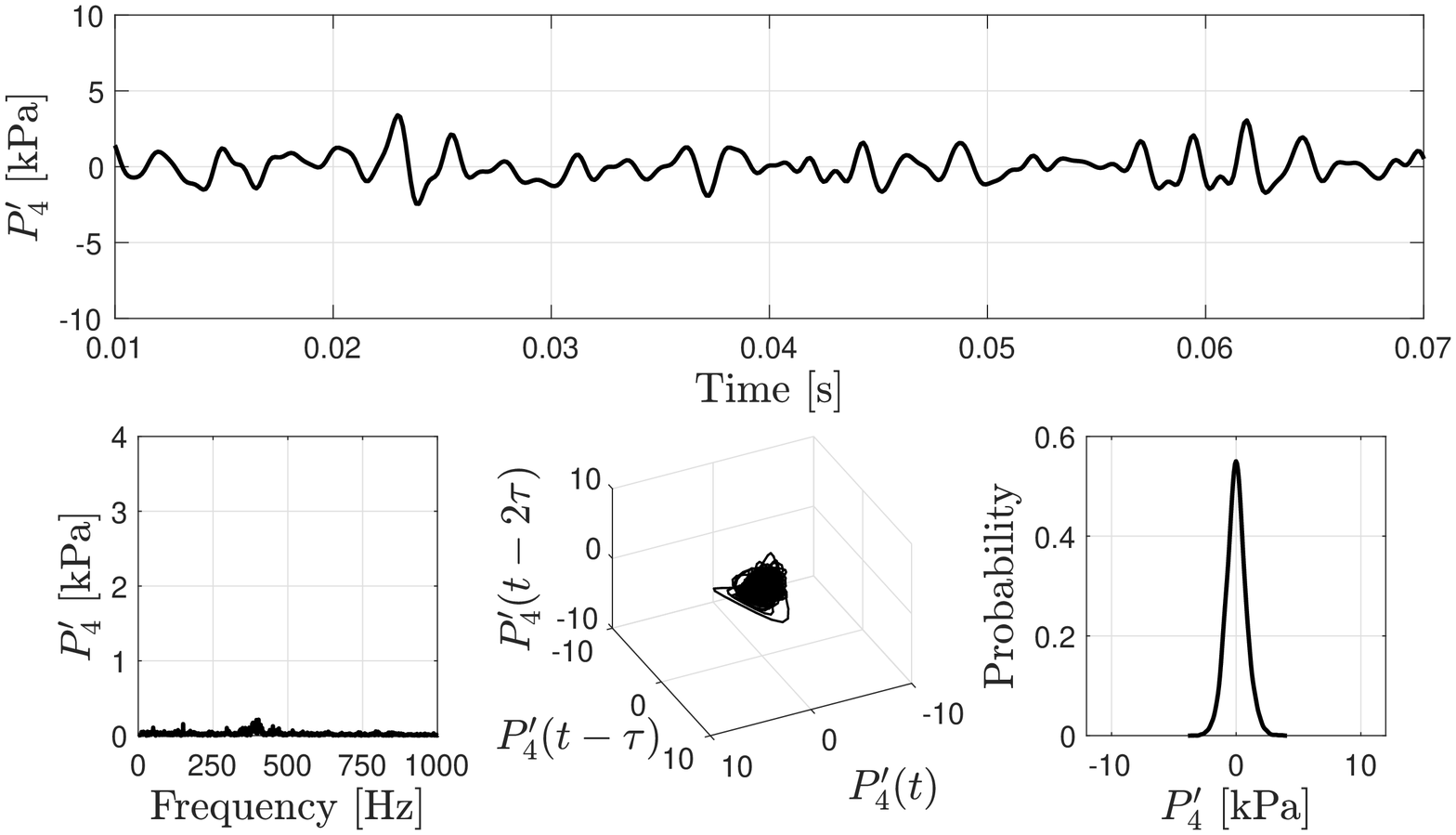}
     }\\
\caption{Pressure fluctuation of P4 sensor for conditions of (a) $T_a=556$ K and (b) $T_a=573$ K (marked as $A$ and $B$ in Fig.~\ref{50}). In each subfigure, the time series are shown in the top half while the spectrum (left), phase space reconstruction (middle), and probability distribution (right) are shown below.}
\label{limitcycle} 
\end{figure}

\begin{figure}[htbp]
\centering

     \subfigure[\label{CH}]{%
       \includegraphics[width=0.35\textwidth]{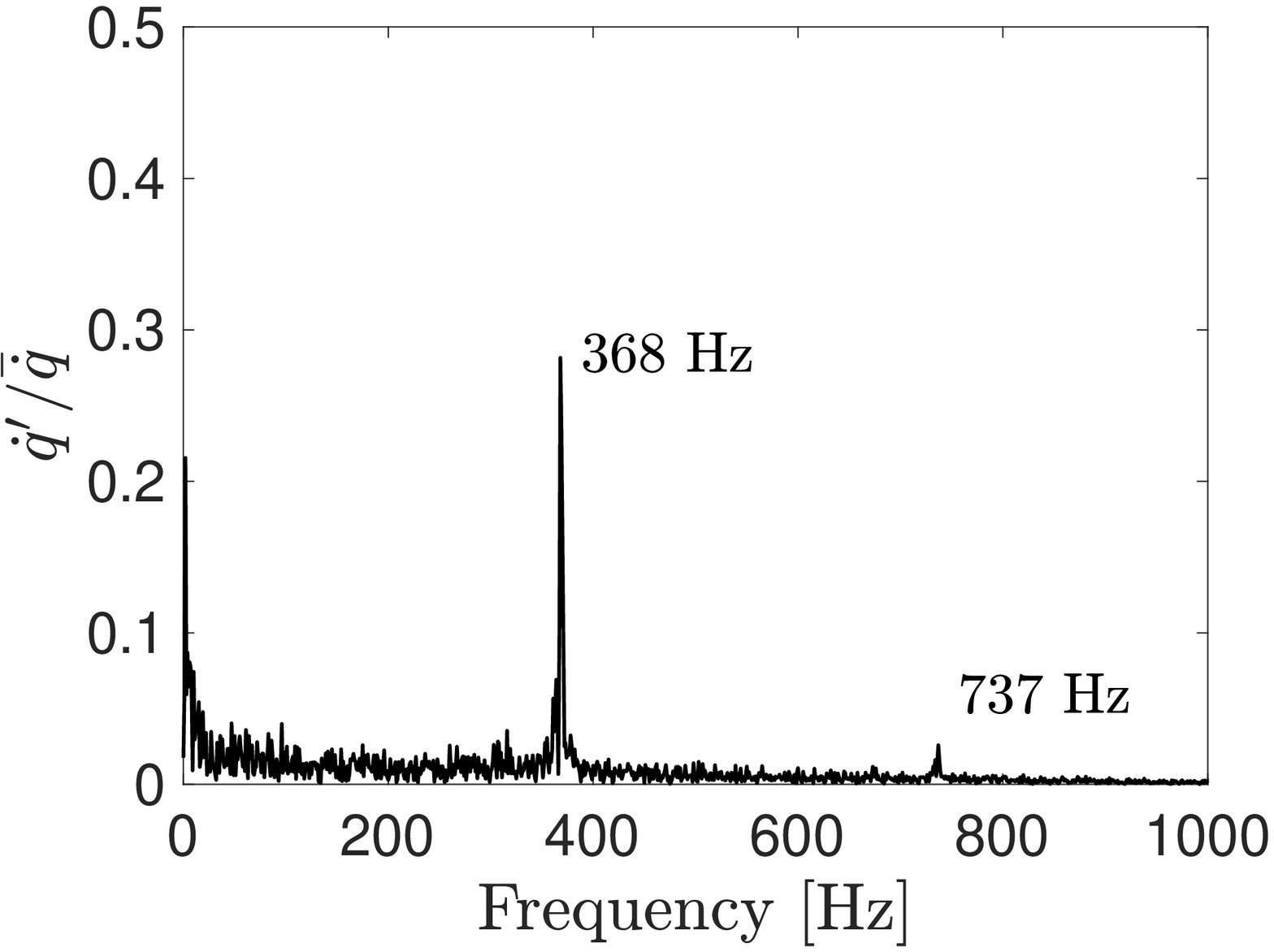}
     }
     \subfigure[\label{FAR}]{%
       \includegraphics[width=0.35\textwidth]{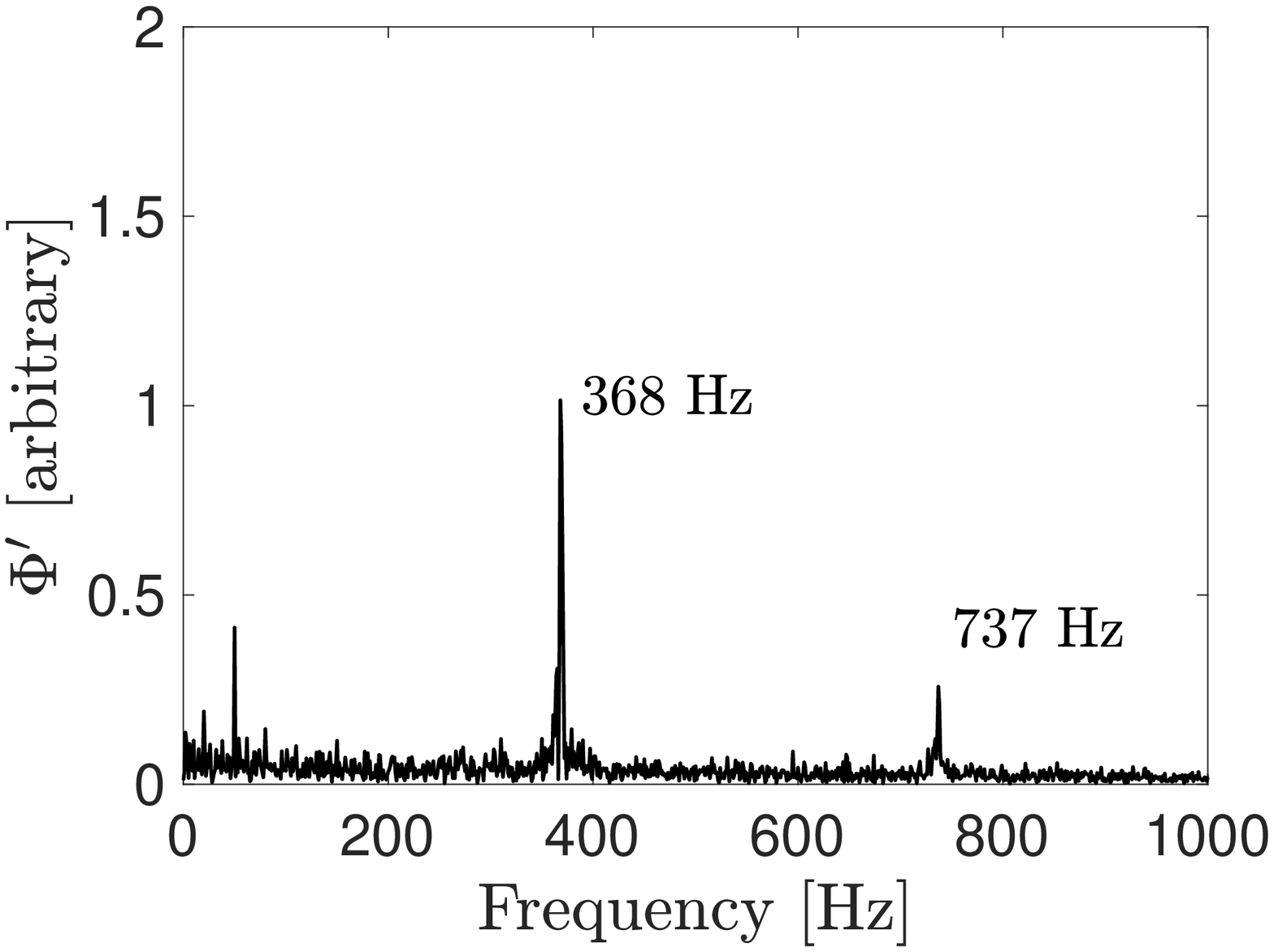}
     }\\
     \subfigure[\label{Pf1}]{%
       \includegraphics[width=0.35\textwidth]{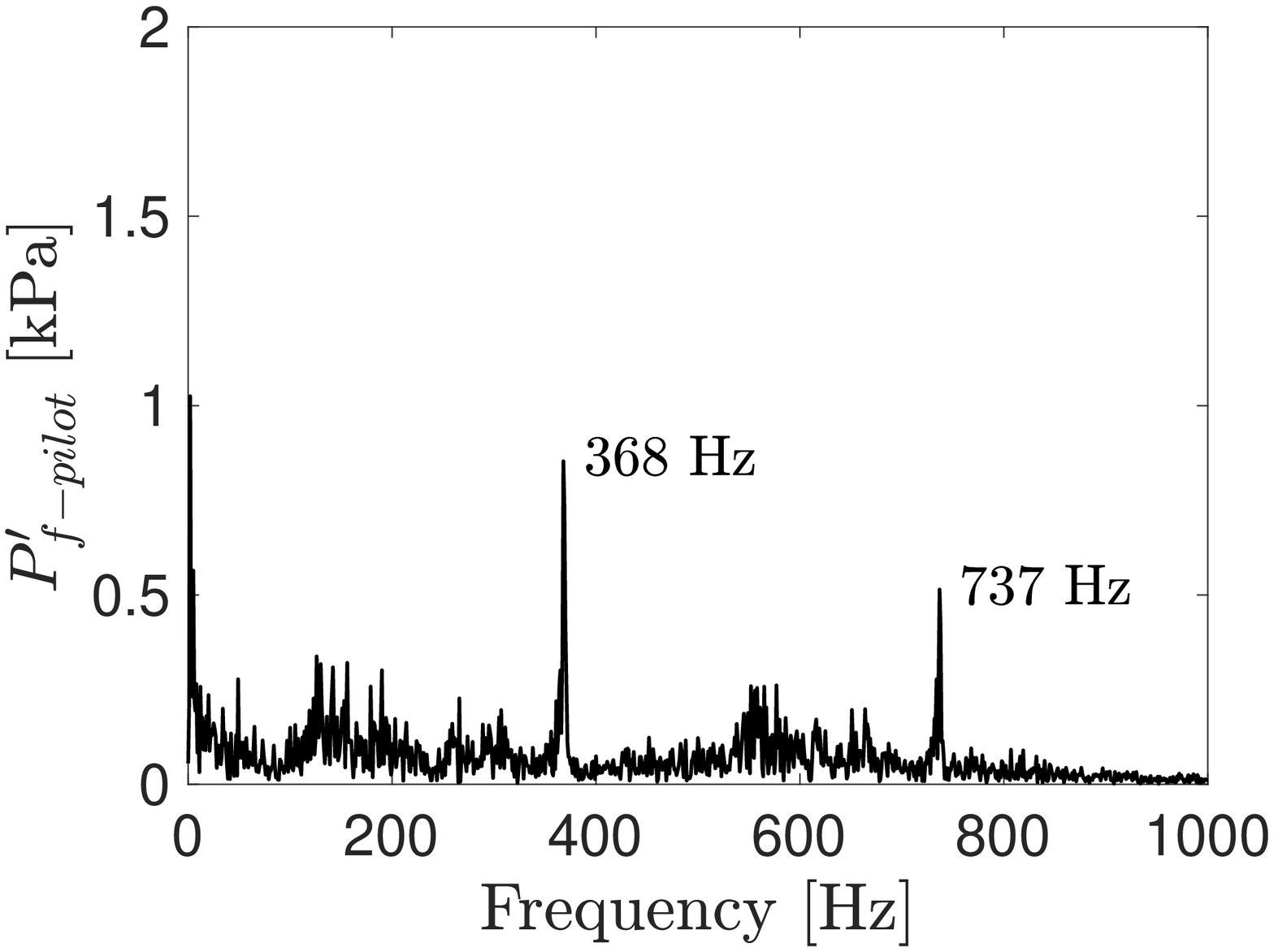}
     }
     \subfigure[\label{Pf2}]{%
       \includegraphics[width=0.35\textwidth]{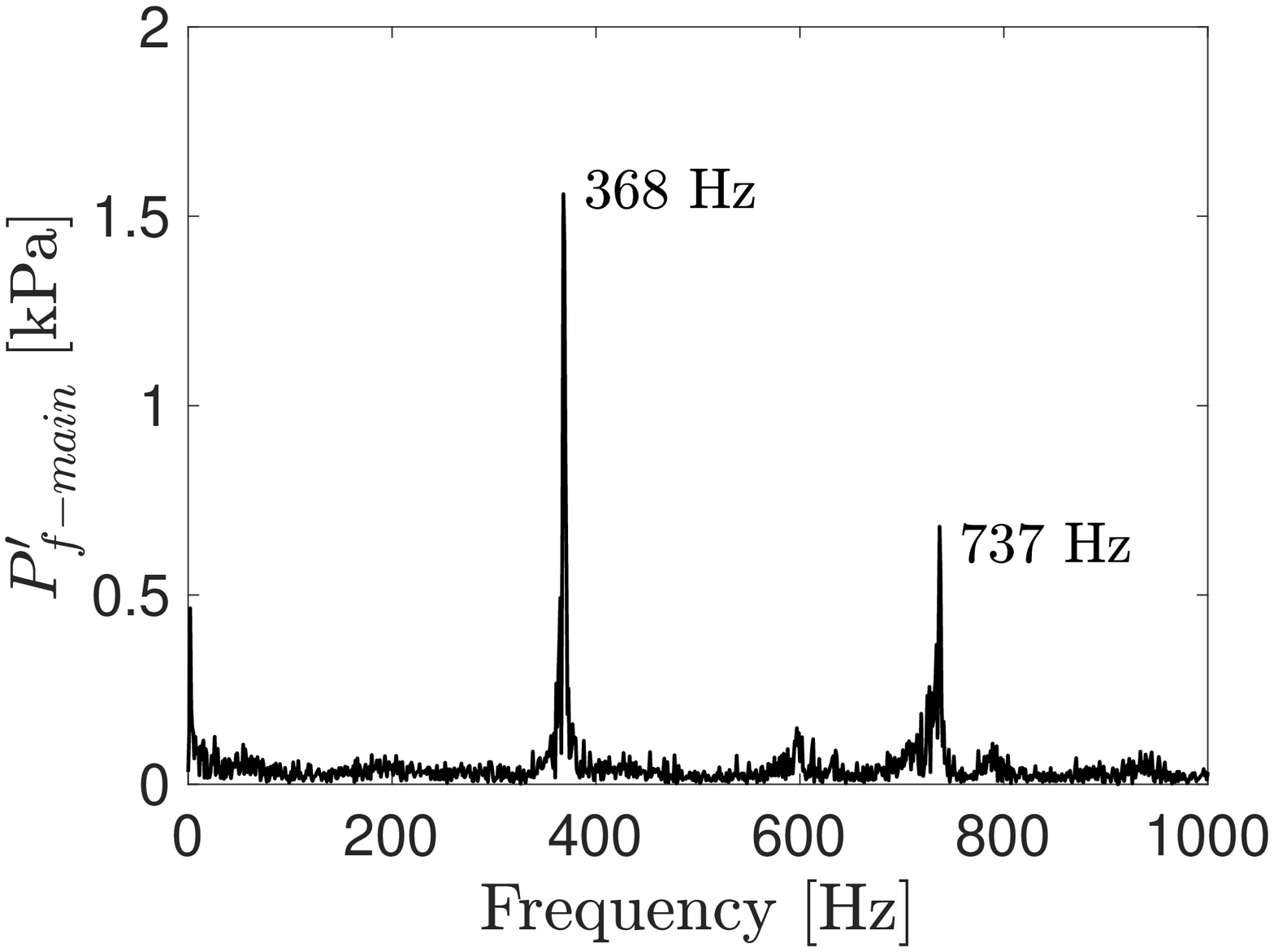}
     }\\
     \subfigure[\label{Mf1}]{%
       \includegraphics[width=0.35\textwidth]{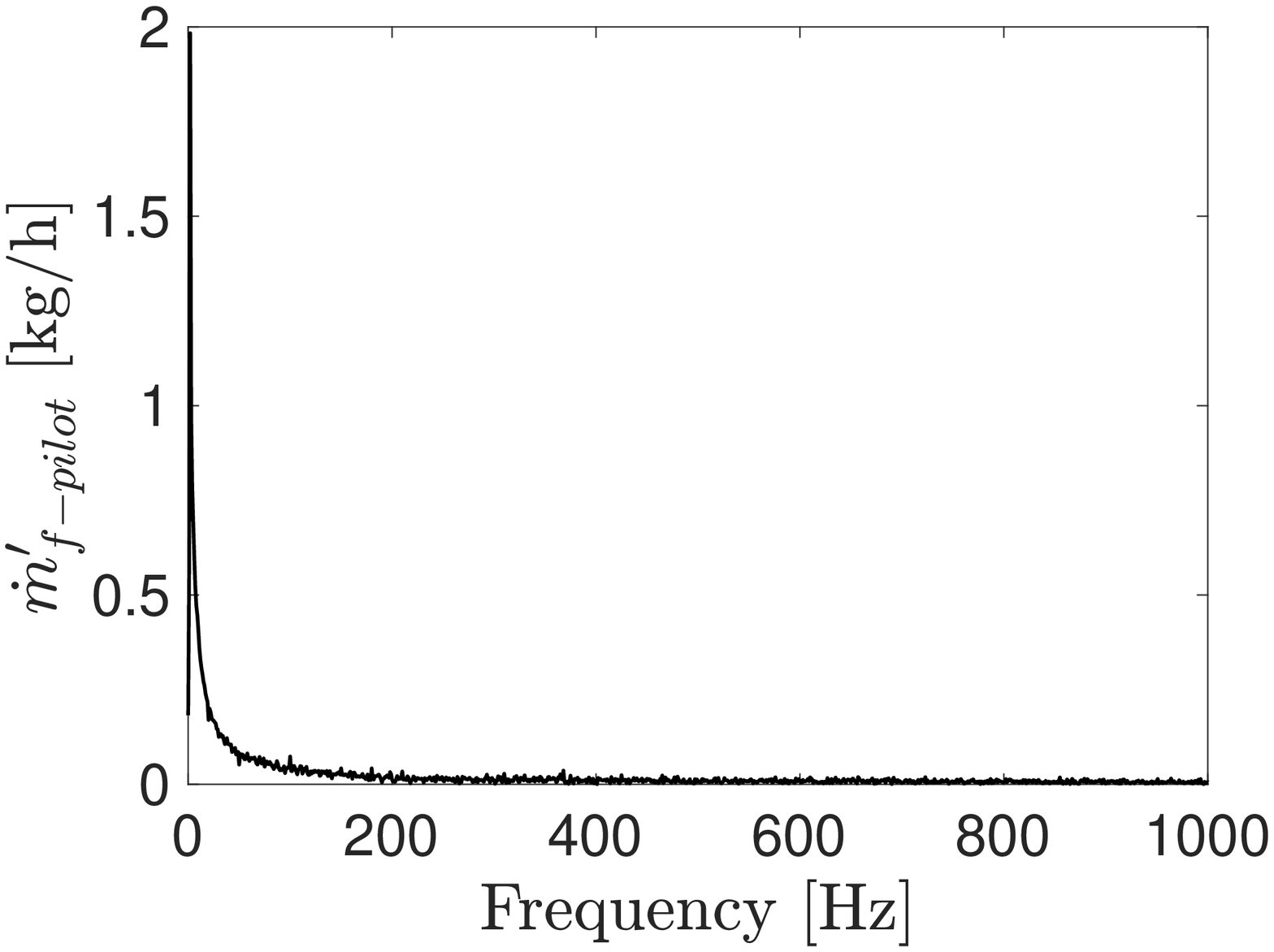}
     }
     \subfigure[\label{Mf2}]{%
     \includegraphics[width=0.35\textwidth]{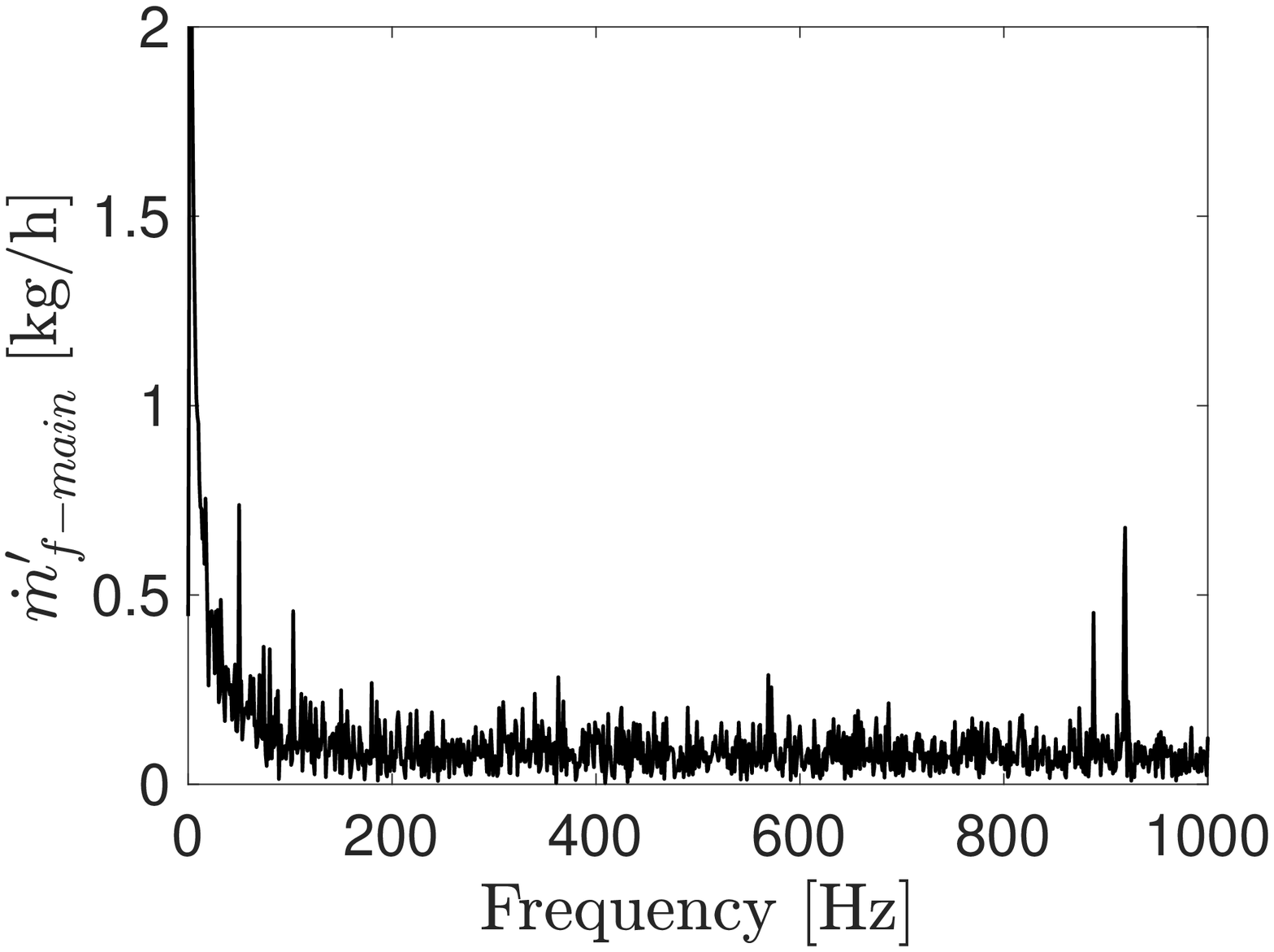}
     }\\
\caption{Spectra of the condition with $T_a=556$ K. (a) normalised heat release rate fluctuation $\dot q'/\bar{\dot q}$ and (b) equivalence ratio fluctuation $\Phi'$. Pressure fluctuation of fuel supply lines of (c) the pilot stage and (d) the main stage. Mass flow rate fluctuation of fuel supply lines of (e) the pilot stage and (f) the main stage.}
\label{spectra1} 
\end{figure}

Figure~\ref{limitcycle}(a-b) reports a further analysis of the pressure signals recorded by the P4 sensor for condition $A$ and $B$. In each subfigure, the time series are shown in the top half while the spectrum (left), phase space reconstruction (middle), and probability distribution (right). The phase space is reconstructed following the method proposed in~\cite{abarbanel1993analysis, guan2018nonlinear}. Firstly, an embedding dimension order of 3 is set for this study, trying to avoid the trajectories collapsing to show them clearly. Secondly, the time lag $\tau_{lc}$ is calculated to shift the signals apart. The auto-correlation function of the original signal ${P_4}'$ is pre-plotted and then the time where the auto-correlation value drops to zero for the first time is chosen as $\tau_{lc}$. Different time lags of $\tau_{lc} = 6/f_s\approx0.73$ ms and $\tau_{lc} = 7/f_s\approx0.85$ ms are chosen conditions $A$ and $B$, respectively. The differences are clear. With $T_a=556$ K, a periodic signal is found with a main frequency 368 Hz and harmonic frequency 737 Hz. A clear annulus trajectory with some noise (caused by the non-linearity of the flame) is also found in the phase space plot, indicating a strong limit cycle type oscillation of the combustion system. A dual-peak probability distribution with wide pressure fluctuation range is found for condition $A$. On the contrary, the pressure signal in condition $B$ features white noise characteristics with no discrete frequency peak. The trajectory in the phase space also collapses to a stable point. The distribution of ${P_4}'$ in condition $B$ follows a typical Gaussian distribution (normal distribution). These features indicate a combustion system that is thermoacoustically stable.

Spectra of heat release rate fluctuations $\dot q'/\bar{\dot q}$ and of equivalence rate oscillations $\phi'$ recorded at the condition $A$ are shown in Fig.~\ref{spectra1} (a) and (b). Both plots are dominated by the same oscillation with a peak frequency at 368 Hz. As observed for the pressure signal ${P_4}'$ (Fig.~\ref{473}), the spectrum of the heat release rate oscillations signal (Fig.~\ref{CH}) shows a harmonic peak at 737 Hz. This second peak is also found in the spectrum of $\phi'$ (Fig.~\ref{FAR}). Spectra of the pressure recordings in fuel supply lines shown in Fig.~\ref{spectra1} (c-f) revealing that the oscillations in both the pilot and the main stage (Fig.~\ref{Pf1} and \ref{Pf2}) are consistent with the one in the main chamber (Fig.~\ref{473}). In contrast, no peak is found in the spectra of the fuel mass flow rates shown in Fig.~\ref{Mf1} and \ref{Mf2}. The pressure drop of liquid fuel in the pilot nozzle and the main stage injectors are of the order of 2.5 bar and 0.5 bar level, respectively, whereas the pressure fluctuations in the fuel supply are of merely 1 kPa level, sufficient to acoustically decouple the two systems. The fuel mass flow rate could then be considered constant in both stages. As a consequence, it is possible to conclude that the equivalence ratio oscillations $\phi'$ found in Fig.~\ref{FAR} are mainly due to the velocity fluctuation of the inlet air. 

\subsection{Flame dynamics}

\begin{figure}[h!]
\centering
\includegraphics[width=0.7\textwidth]{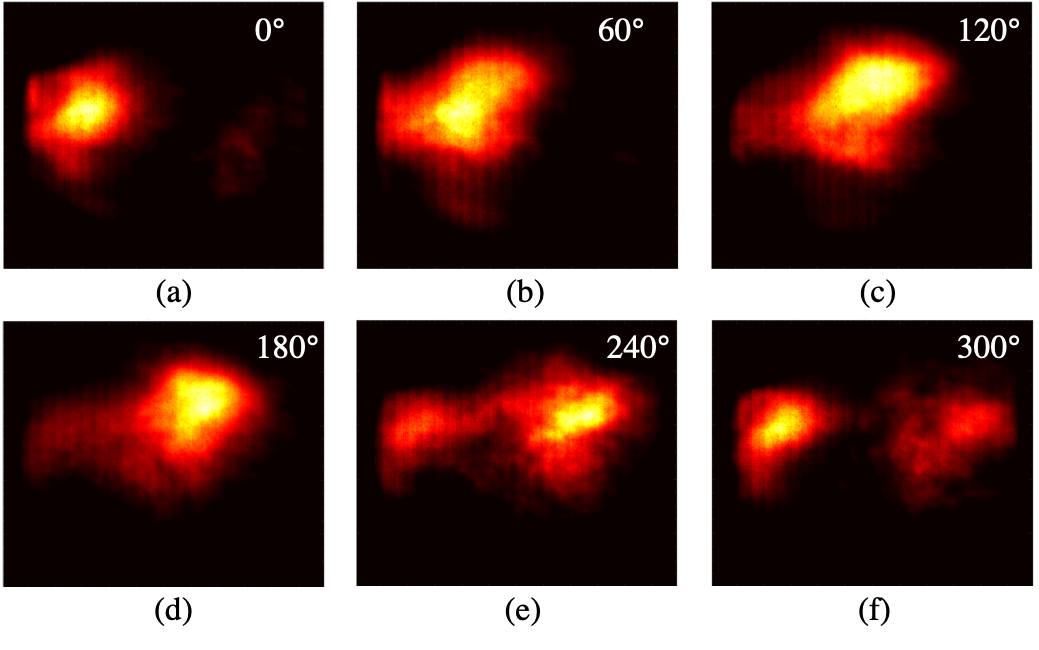}
\caption{Phase-averaged flame snapshots of the case with an inlet temperature of 556 K (condition $A$). Flow from the left to the right.}
\label{FlamePhz}
\end{figure}

The CH* chemiluminescence of flame is captured by the intensified high-speed camera with a CH* filter. The flame dynamics recorded under the unstable case for $T_{a}$=556 K (condition $A$ in Fig.~\ref{50}) are now discussed. Figure~\ref{FlamePhz} reports flame snapshots at six different times during the instability cycles. Experimental images are phase-averaged over 60 cycles. As already experienced in previous studies of spray flames~\cite{temme2014combustion, ahn2018low}, a periodic axial movement with flame lift-off and re-ignition is clearly observed: (1) at $\theta$=0$^\circ$ (Fig.~\ref{FlamePhz}(a)) the flame appears attached to the injector; (2) Then it moves downstream with a subsequential lift-off from the injector at $\theta$=60$^\circ$-180$^\circ$ (Fig.~\ref{FlamePhz}(b-d)). (3) Finally, at $\theta$=240$^\circ$(Fig.~\ref{FlamePhz}(e)) the flame re-ignites downstream of the injector. At $\theta$=300$^\circ$(Fig.~\ref{FlamePhz}(f)) the new flame kernel expands while the lifted flame tends to disappear. 

\begin{figure}[htbp]
\centering
\includegraphics[width=0.7\textwidth]{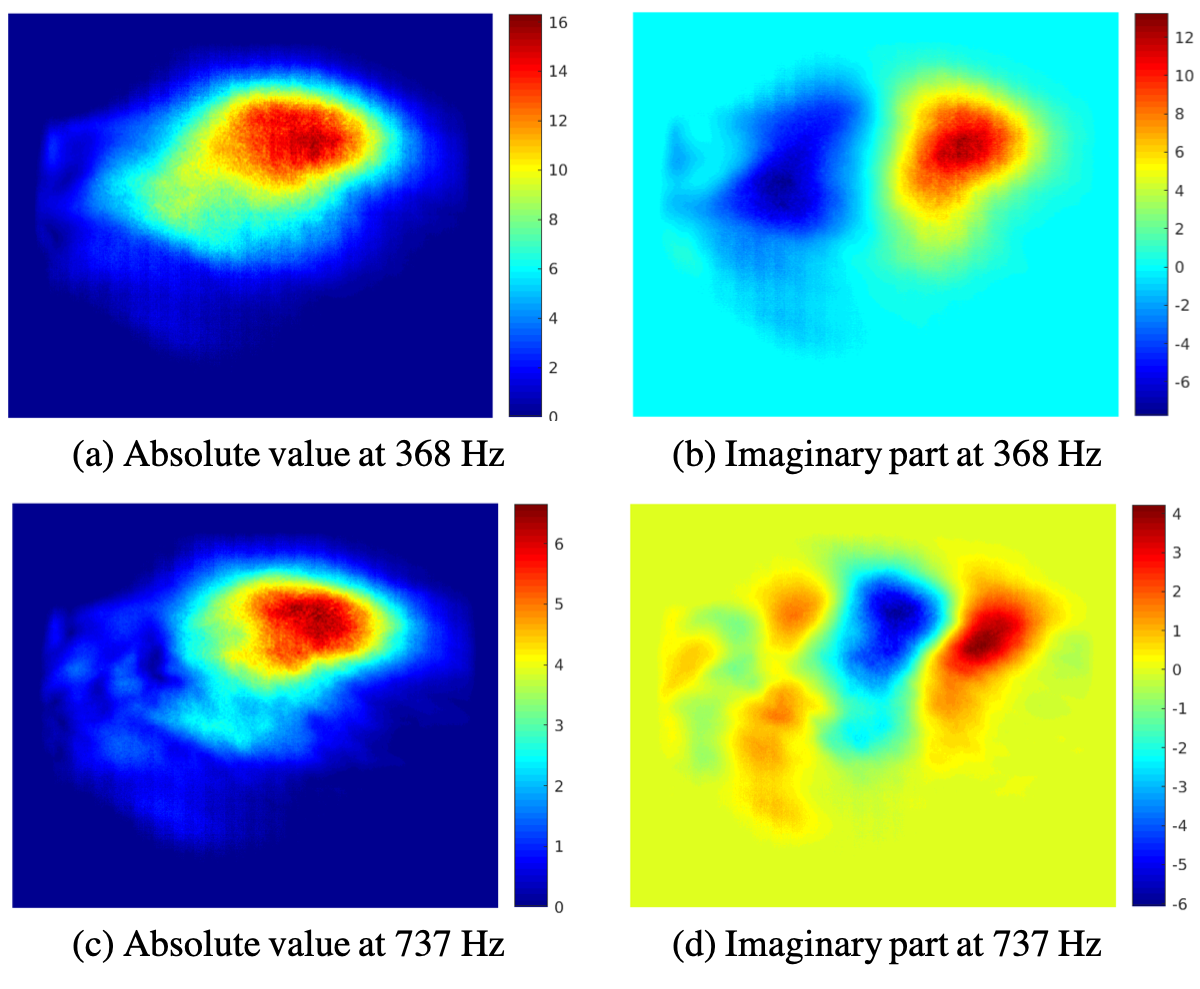}
\caption{The pixel-to-pixel Fourier transform of flame images of the case with the inlet temperature of 556 K. (a) Normalised absolute value and (b) normalized imaginary part of flame dynamics at domain frequency at 368 Hz. (c) Normalised absolute value and (d) normalised imaginary part of flame dynamics at a harmonic frequency of 737 Hz.}
\label{FlameFFT}
\end{figure}

To better highlight the flame coherent structures at specific frequencies, a pixel-to-pixel Fourier transform method is performed on the flame snapshots. The dynamics of the mode at 368 Hz (the main frequency) and 737 Hz (the first harmonic) are shown in Fig.~\ref{FlameFFT}: the absolute values of the FFT results (Figs.~\ref{FlameFFT}(a,c))  indicate the global structures of the flame mode. The coherent flame structures are better shown by the normalised imaginary part of the two frequency shown in Figs.~\ref{FlameFFT} (b,d). Both representations clearly confirm the axial periodic movement of the flame. The difference is that the mode in Fig.~\ref{FlameFFT} (b) ($f$=368 Hz) is characterised by a length $\lambda$=100 mm, twice than that obtained in Fig.~\ref{FlameFFT} (d) ($f$=737 Hz).

\begin{figure}[htbp]
\centering
     \subfigure[\label{530}]{%
       \includegraphics[width=0.6\textwidth]{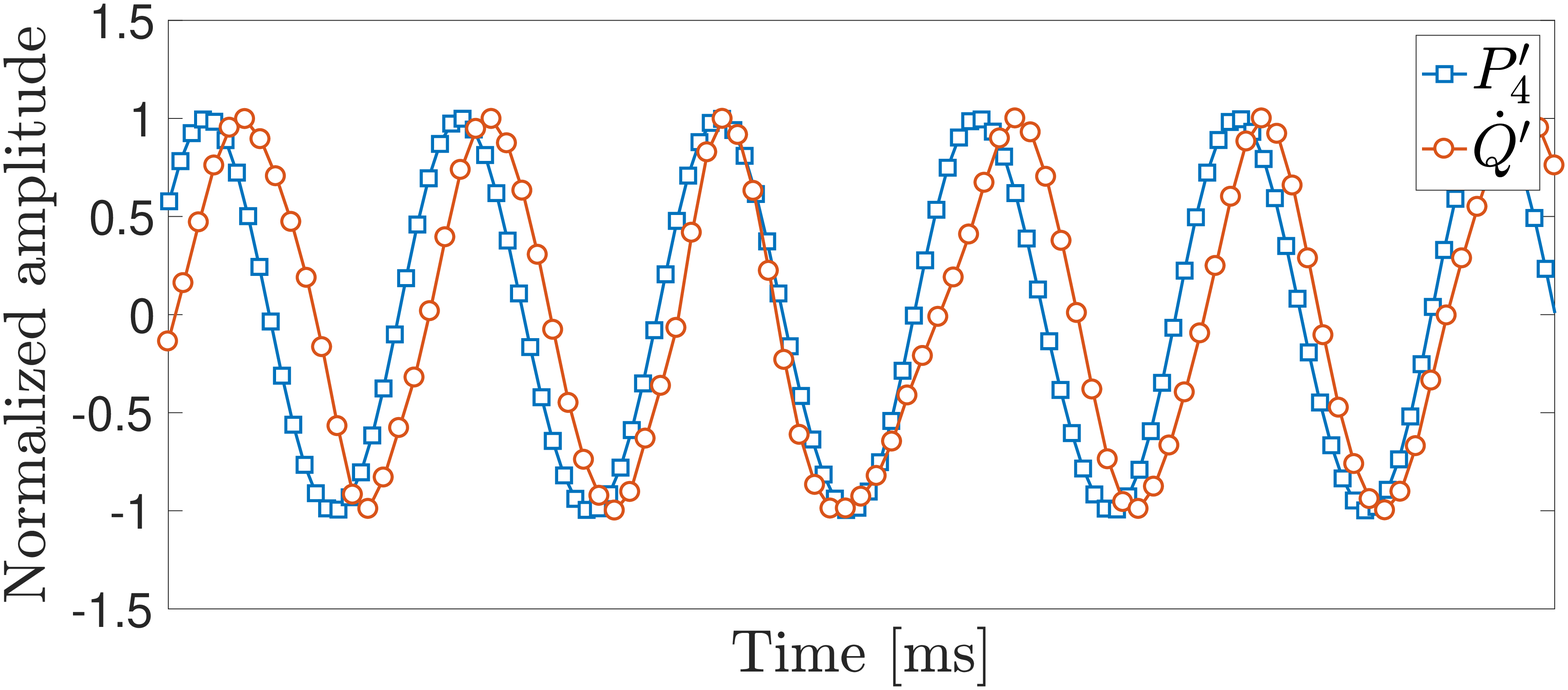}
     }\\
     \subfigure[\label{556}]{%
       \includegraphics[width=0.6\textwidth]{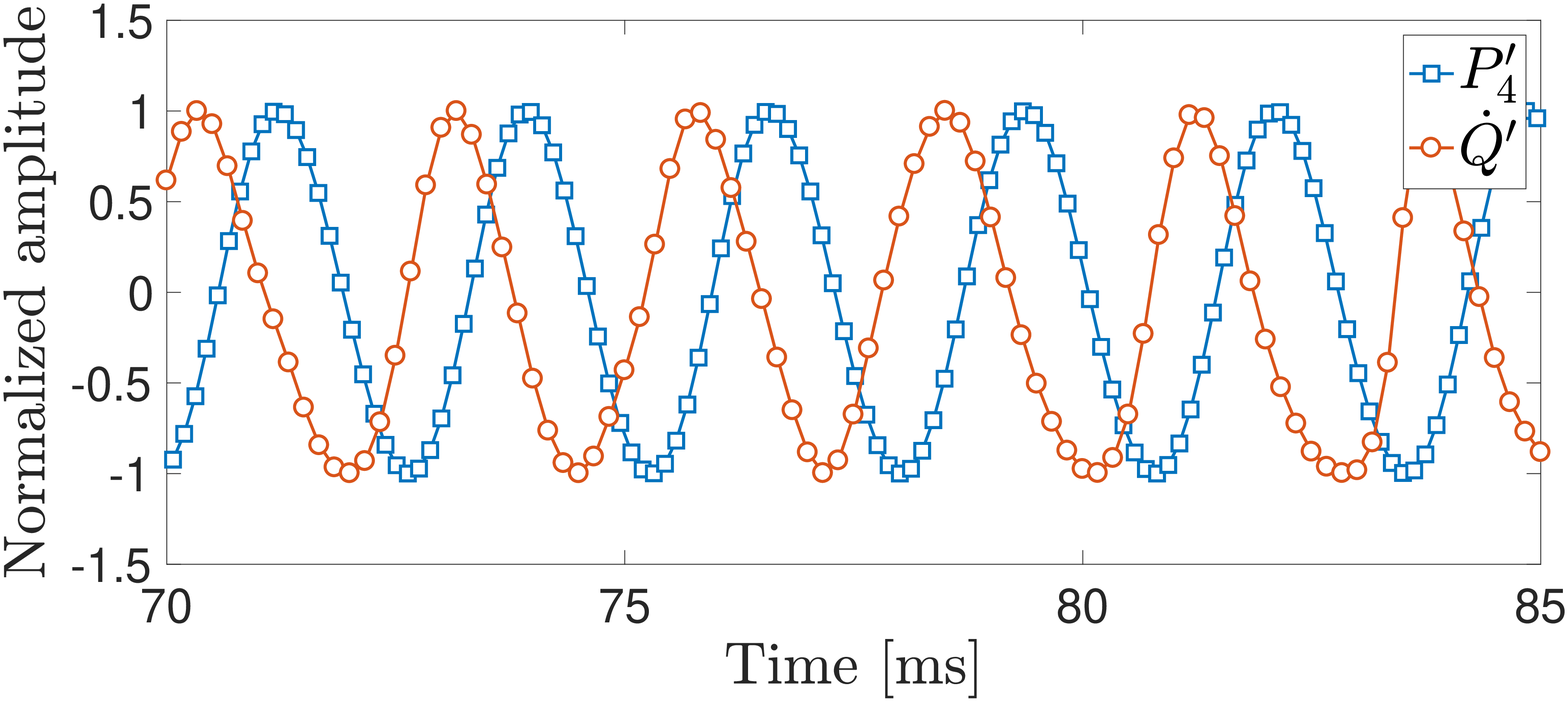}
     }\\
\caption{Phase relationship of ${P_4}'$ and intensity fluctuation of flame images $\dot{Q}'$. Unstable cases: (a) $T_a=530$ K, (b) $T_a=556$ K.} 
\label{phaseshift} 
\end{figure}

Finally, space-integrating the intensity of the flame images, the time-signal of the total heat release rate $\dot{Q}'$ can be retrieved. The signals of ${P_4}'$ and ${\dot{Q}}'$ are normalized by Hilbert transform and plotted in Fig.~\ref{phaseshift}. Figure ~\ref{phaseshift} allows a clear visualisation of the phase relationship between the two signals, which according to the Rayleigh criterion~\cite{rayleigh1878explanation} regulates the positive feedback between $p'$ and $\dot q'$ and, therefore, the entire thermoacoustic stability properties of the system. In the first unstable point of the bifurcation map, i.e., $T_{a}$=530 K shown Fig.~\ref{530}, heat release and pressure fluctuations are almost synchronous, with ${\dot{Q}}'$ being ahead of ${P_4}'$. Indeed for this point, the maximum oscillation amplitude is obtained. With the increase of inlet temperature, for the $T_{a}$=556 K case, the phase difference between the two signals increases (with ${\dot{Q}}'$ being behind ${P_4}'$), merely meeting the Rayleigh criterion. This change of phase difference is possibly caused by the change of convective time $\tau$ of perturbation passing through the flame. As $T_{a}$ rises, $\tau$ decreases. The decreasing of $\tau$ leads to a different thermoacoustic coupling, resulting in enlarged phase shift between ${P_4}'$ and ${\dot{Q}}'$. This will be further discussed in next section. For $T_{a}=573$ K, ${\dot{Q}}'$ and ${P_4}'$ are out of phase and the constructive coupling of ${P_4}'$ and ${\dot{Q}}'$ is ruined. The system then stabilises with no thermoacoustic instability. The signal of $\dot{Q}'$ is dominated by noise and is therefore not shown.

\section{Discussion and analysis}

A theoretical study is now proposed to explain the previously discussed bifurcation phenomenon.
\subsection{Thermo-acoustic analysis }
To predict thermoacoustic instability, a model for the acoustic waves (their generation by the flame unsteadiness, their propagation  within the combustor and their reflection from boundaries) needs to be combined with a  model for how the unsteady heat release rate of the flame responds to acoustic disturbances~\cite{lieuwen2005combustion,dowling2003acoustic,schuller2003unified}. In this work, following Crocco and Cheng~\cite{crocco1956theory} a time delay n--$\tau$ model is assumed. In the time domain this model is defined as
\begin{equation}
 \frac{\dot q'(t)}{\overline{\dot q}}=n\frac{u_i'(t-\tau)}{\overline{u}_i},
\label{eq:n-t}
\end{equation}
where $n$ is defined as the acoustic-combustion interaction index that controls the amplitude of the flame response, $u_i$ is the velocity fluctuation, and $\tau$ is the time delay between the acoustic perturbations and the flame response. Depending on the combustion system, different physical mechanisms are responsible for the fluctuations of heat release oscillations and each one of them is characterized by a different characteristic timescale~\cite{lieuwen2005combustion}.  One strength of the time delay model proposed by Crocco is that the value assumed for $\tau$ can be chosen to describe the physical influence of the specific driving mechanism. This work is focused on combustion instabilities occurring in gas turbine combustors~\cite{hermeth2014bistable}. Experimental tests~\cite{lieuwen1998role} show that for these flames the main cause of heat release fluctuations ($\dot q'$) is due to oscillations of the equivalence ratio $\phi'$ which are caused by fluctuations of the air flow rate at the fuel injectors. Mathematically, this results in
\begin{equation}
\frac{\dot q'}{\overline{\dot q}}\sim\frac{\phi'_{inj}(t-\tau)}{\overline{\phi}_{inj}},
\label{time_lag_inj}
\end{equation}
where $inj$ refers to the injection location. The equivalence ratio fluctuations at the injection point can be derived from its definition
\begin{equation}\label{phi_fluctu}
\frac{\phi'_{inj}(t-\tau)}{\overline{\phi}_{inj}}\sim\left [\frac{\dot m_f'(t-\tau)}{\overline{\dot m}_f}-\frac{\dot m'_a(t-\tau)}{\overline{\dot m}_a} \right],
\end{equation}
where $\dot m_f$ is the mass flow rate of the fuel and $\dot m_a$ is the mass flow rate of the air, $\dot m_a$=$\rho_{inj}u_{inj}A_{inj}$. Experimentally was observed that the injector is choked (i.e., $\dot m_f'$=0), so the equivalence ratio fluctuations depend only on $\dot m_a'$. Furthermore, neglecting the mean flow it is possible to conclude that equivalence ratio fluctuations, and, due to Eq.~\ref{time_lag_inj}, heat release rate fluctuations are mainly influenced by velocity fluctuations. Following these assumptions, Eq.~\ref{time_lag_inj} combined with Eq.~\ref{phi_fluctu} results
\begin{equation}\label{n_tau_phi}
\frac{\dot q'}{\overline{\dot q}}=-n\frac{u'_{inj}(t-\tau)}{\overline{u}_{inj}}.
\end{equation}

The thermo-acoustic analysis can be carried out in the frequency domain~\cite{lieuwen2005combustion, poinsot2005theoretical, dowling2003acoustic}. Applying the harmonic analysis to Eq.~\ref{n_tau_phi}, in the limit of small perturbations~\cite{dowling2003acoustic}, the linear  Flame Transfer Function (FTF) is derived
\begin{equation}\label{eq:linear_n_tau}
\mathcal{T}_L=\frac{\hat u_{inj}/\overline{u}_{inj}}{\hat{\dot q}/\overline{\dot q}}=-ne^{-i\omega\tau}.
\end{equation}

\begin{figure}[ht]
\centering 
\includegraphics[width=0.5\textwidth]{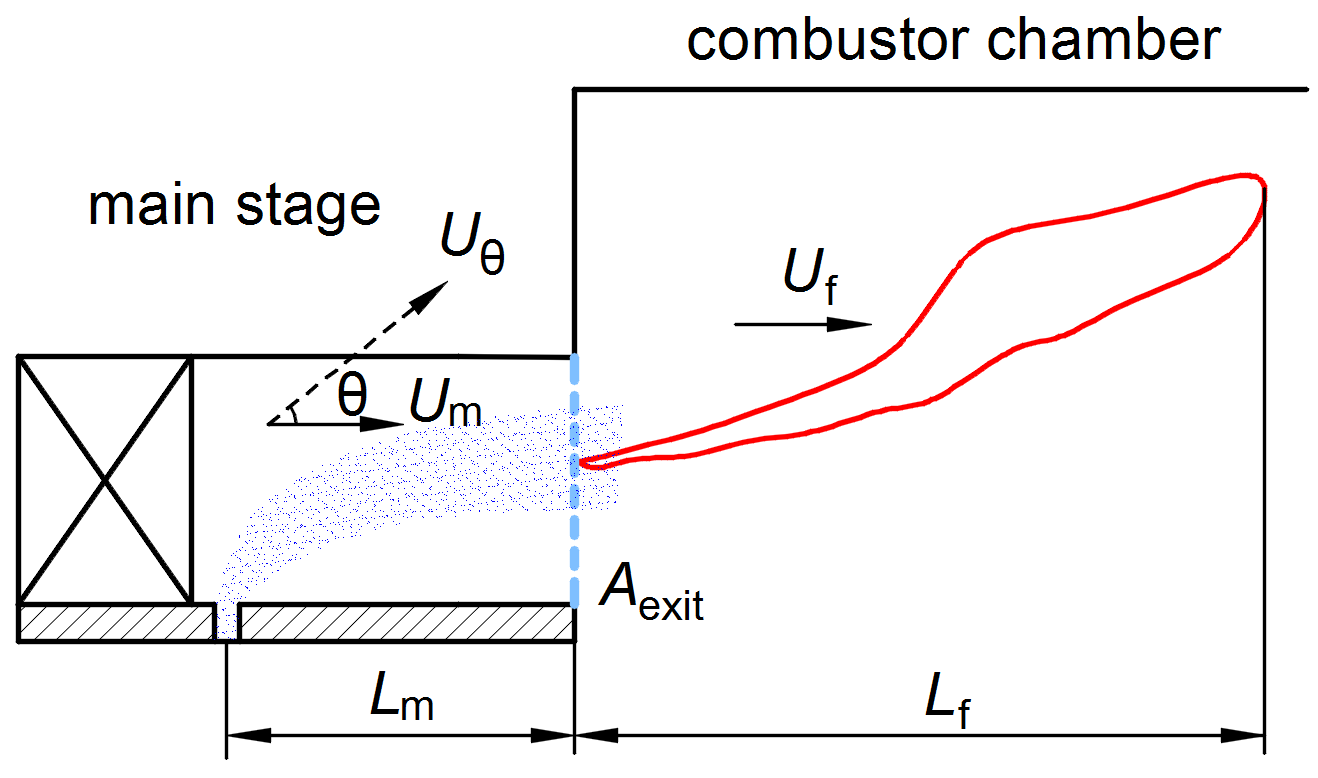}
\caption{Schematic of convection time from the fuel injection to the flame tip.}
\label{spray}
\end{figure}
An analytical approximation of the time delay $\tau$ of Eq.~\ref{eq:linear_n_tau} will be proposed. In this stage, only the main stage of the LPP combustor sketched in Fig.~\ref{spray} will be considered, since it delivers the major part of the air and the fuel in the combustor. Under the hypothesis of neglecting the injector lags and the chemical reaction times~\cite{lieuwen2005combustion}, the time delay $\tau$ correlated to the heat release rate driving mechanism can be assumed to be comprised of two parts:
\begin{equation}
 \tau = \tau _{m}+ \tau _{f}.
\label{tau}
\end{equation}
where $\tau _{m}$ is the mixing time and $\tau _{f}$ is the convection time of perturbation moving from the flame base to the flame tip. Since the majority of the mixing happens in the main stage, it can be estimated as the ratio between the distance between the fuel injector and the exit of the swirlor $L_m$=23 mm (see Fig.~\ref{spray}) over the axial velocity  $U_m$ inside the main stage, i.e., $\tau _{m}=L_m/U_m$ . Following~\cite{lieuwen2001mechanism}, $\tau _{f}$ can be assumed to be computed as $\tau _{f}=L_f/U_f$, where $L_f$ is the length from the flame base to the flame tip, a constant flame length of $80$ mm is chosen. $U_f$ is the axial velocity of the fluid velocity at the flame location. Considering the converging exit of the main stage, it is mainly controlled by the exit area $A_{exit}$ and changes with different inlet temperatures $T_a$.  The axial velocity inside the main stage $U_{m}$ is then estimated as 0.38 times of $U_f$. The estimated convection time for all the inlet air temperature $T_a$ is listed in Tab.~\ref{tab_tau}.
\begin{table}[ht]
\centering
\caption{Estimated convective time of cases with $T_a$ ranging from 530 K to 617 K.}
\label{tab_tau}
\begin{tabular}{cccccc}
\hline
$T_a$ (K)  & $f$ (Hz)  & $\tau_m$ (ms) & $\tau_f$ (ms) & $\tau$ (ms) & Status \\\hline

530   & 358  & 0.87                  & 1.17                  & 2.04   & unstable \\
556 (Case A)   & 368  & 0.83                  & 1.12                  & 1.95 & unstable\\
573 (Case B)   & /  & 0.81                  & 1.08                  & 1.89  & stable\\
589   & /  & 0.79                  & 1.05                 & 1.84                  & stable\\
596   & /  & 0.78                  & 1.04                 & 1.82                  & stable\\
617   & /  & 0.75                 & 1.01                 & 1.76                  & stable\\\hline

\end{tabular}
\end{table}

%%%%%%%%%%%%%%%%%%%%%%%%%%%%%%%%%%%%%%%%%%%%%%%%%%
\begin{figure}[htbp]
\centering
\includegraphics[width=0.7\textwidth]{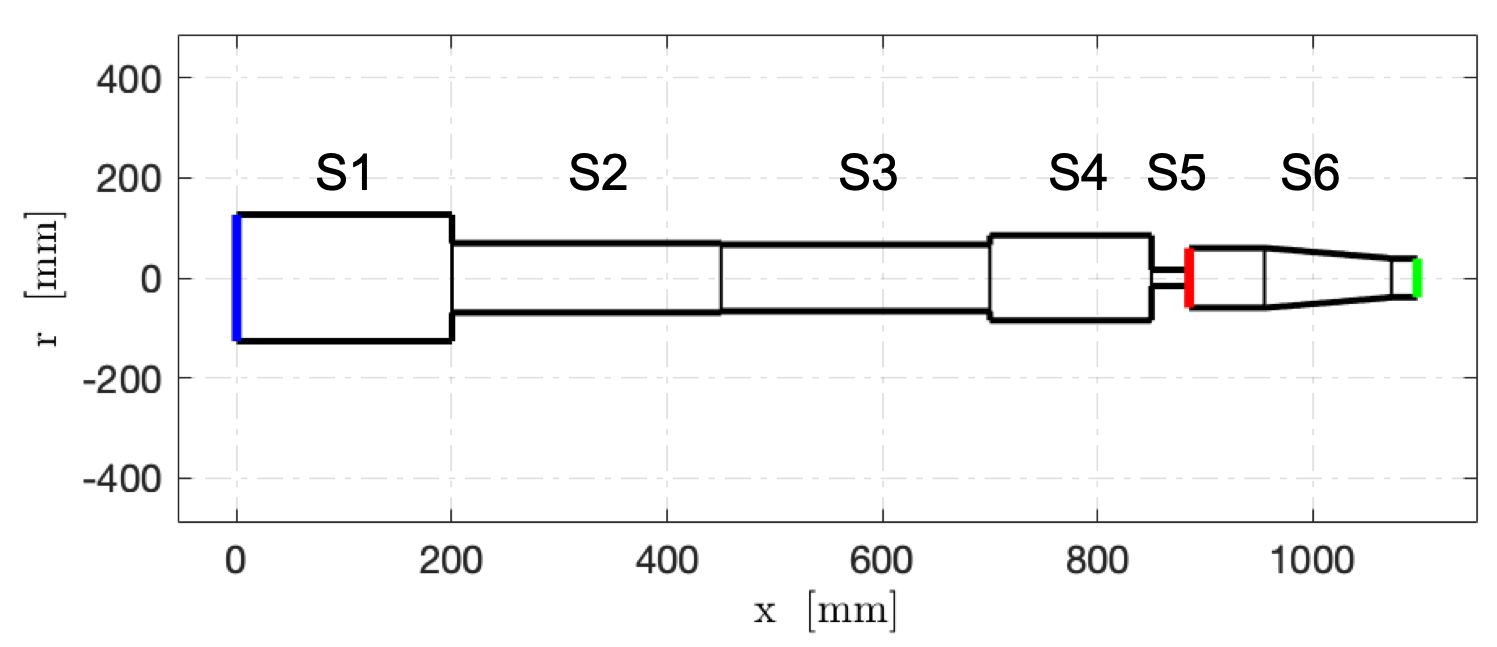}
\caption{Simplified thermoacoustic network geometry used in OSCILOS.}
\label{geo} 
\end{figure}

%%%%%%%%%%%%%%%%%%%%%%%%%%%%%%%%%%%%%%%%%%%%%%%%%%

The defined FTF is now coupled in the open source low order network solver OSCILOS\footnote{http://www.oscilos.com}~\cite{li2015time,yang2019systematic} to predict the instability of the system for the different operative conditions. This code has been widely used for thermo-acoustic analyses and validated, e.g., see Refs.~\cite{han2015openfoam, li2017numerical} for atmospheric condition and ~\cite{xia2019numerical} for high pressure configuration, and ~\cite{yang2019systematic} for annular combustors. It represents the combustor geometry as a network of connected simple modules, as shown in Fig.~\ref{geo}.  The length and cross-sectional area of each module reported in Tab.~\ref{tab_geo} match the original geometry. The mean flow is accounted for, with the mean flow variables assumed constant within each module, changing only between modules.

\begin{table}[ht]
\centering
\caption{Geometry parameters for low order thermoacoustic model analysis.}
\label{tab_geo}
\begin{tabular}{cccc}
\hline
Section  & Name                 & Area (mm2) & Length (mm) \\ \hline
S1   &  Heat exchanger plenum      & 50247      & 200         \\
S2   &  Measuring section                   & 15175      & 250         \\
S3  &  Flow straightening  section & 13860      & 250         \\
S4   &  Combustor plenum         & 22776      & 150         \\
S5  &  Swirler        & 820        & 35          \\
\multirow{3}{*}{S6}  &  Flame liner 1              & 11235      & 70          \\
  &   Flame liner 2              & 11235$\rightarrow$4708      & 118         \\
   &  Flame liner 3              & 4708       & 24         \\ \hline
\end{tabular}
\end{table}

As it is common practice in low-frequency thermo-acoustic analyses, that the acoustic waves are considered linear and one-dimensional. Thus within each module acoustic perturbations can be represented as the sum of upstream and downstream travelling waves with different strengths. Linearised flow conservation equations are assumed between modules, accounting for losses due to the stagnation pressure drop at area expansions~\cite{davies1988practical}. A pressure node is assumed at the plenum inlet (indicated in blue in Fig.~\ref{geo}), whereas the combustion chamber exit ({indicated as choked orifice in Fig.~\ref{schematic}) is approximated as choked, by imposing an acoustic velocity node~\cite{marble1977acoustic}. Given the compactness of the flame axial extension with respect to the wavelength of the measured acoustic modes, the heat release zone is assumed concentrated in an infinitely "thin" flame sheet domain (indicated in red in Fig.~\ref{geo}). The linearised flow conservation jump equations proposed by ~\cite{dowling2003acoustic} are imposed to link the acoustic strengths across this domain. 

\begin{figure}[htbp]
\centering
\includegraphics[width=0.65\textwidth]{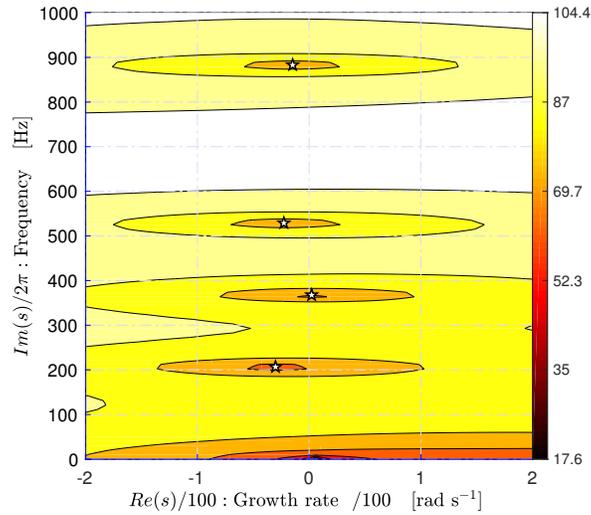}
\caption{Stability map of $T_a=556$ K (Case A), modelling with n=1.85 and $\tau=1.9$ ms. }
\label{stabilitymap} 
\end{figure}

At first, the complete stability map of the system is computed considering an inlet air temperature of $T_{a}$=556 K for which a time delay of $\tau$=1.95 ms was estimated (Tab.~\ref{tau}), as shown in Fig.~\ref{stabilitymap}. In this analysis, a constant gain of n=1.85 is assumed. The complex frequencies, $\omega$ = $\sigma$ + $i2\pi f$ (with $\sigma$ the growth rate), for which both the inlet and outlet boundary conditions are satisfied, are identified within OSCILOS. The computed modes are identified with white stars, showing that a mode at a frequency $f=367.7$ Hz, close to the one measured experimentally (368 Hz), is predicted to be unstable (with positive growth rate but very close to zero). In Fig.\ref{modeshape} the numerical mode shape (continuous blue line) of the unstable mode is compared with the experimental measurements (rectangular marks) from the six sensors P1 to P6 along the test rig. The acoustic pressure is normalised with the corresponding amplitude of the P4 sensor. In agreement with the experimental measurements, the pressure amplitude remains low in the plenum and features a jump in the combustion chamber where remains almost constant until the exit. This implies that the combustor chamber works as the volume of the Helmholtz resonator. This behaviour suggests that the combustion instabilities are coupled with the Helmholtz mode or bulk mode of the system, which is a behaviour that has been previously experienced in combustors featuring liquid fuel combustion~\cite{temme2014combustion, ahn2018low}.

\begin{figure}[htbp]
\centering
\includegraphics[width=0.55\textwidth]{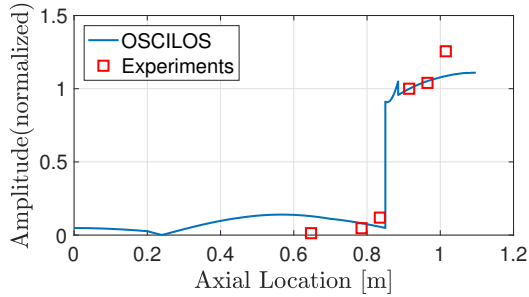}
\caption{Comparision of mode shape in the condition with $T_a=556$ K (Case A).}
\label{modeshape} 
\end{figure}

The impact of the inlet air temperature on the stability of the combustor is now investigated. The previously described thermoacoustic analysis is repeated changing the inlet air temperature $T_a$ and, accordingly, the FTF time delays $\tau$. The analysis is repeated for three different values of the FTF gain, ranging from 1.7 to 2. Differently from before, only the unstable mode around 350 Hz of the system is considered. Results are shown in Fig.~\ref{oscilos} where the dependances of the resonant frequency $f$ and growth rate  $\sigma$ on the FTF gain $n$ and time delay $\tau$ is shown. Regardless of the value assumed for the interaction index $n$, increasing the inlet air temperature, i.e., decreasing the time delay, results in an increase of the resonant frequency and a reduction of the growth rate. This trend is in line with the experimental results presented in the previous sections, in which the system features less intense oscillation amplitudes for increasing inlet temperature until a stable condition is reached. Unfortunately, limit cycle amplitude information cannot be captured by the proposed linear analysis, this requires a nonlinear flame model~\cite{dowling1997nonlinear,laera2017weakly}. Nevertheless, a more solid validation of the proposed model can be achieved by focusing the attention on the condition for which a stability shift is predicted. As highlighted in Fig.~\ref{oscilos}, the system is predicted to be unstable until a time delay of $\tau\approx 1.9$ ms and a $T_a\approx 556$ K are imposed in the model. This is in good agreement with the conditions observed at the Hopf bifurcation in both reported datasets (marked as $A$ in Fig.~\ref{50}). With higher $n$ (higher flame response to acoustic perturbation), the growth rate lines move upward, meaning the system is moving towards to the unstable region, confirming the robustness of the proposed model.

\begin{figure}[htbp]
\centering
\includegraphics[width=0.5\textwidth]{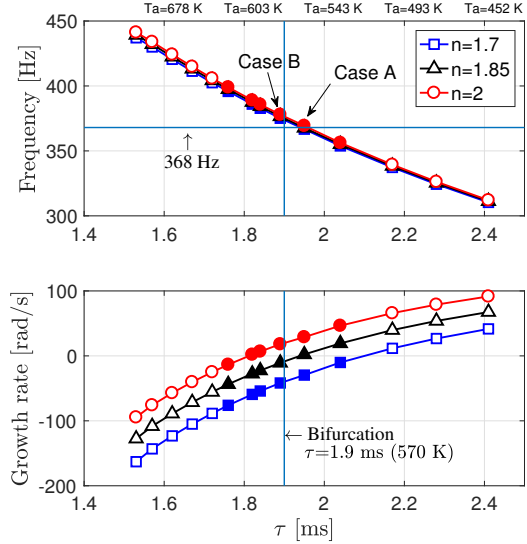}
\caption{Dependence of the predicted frequency and growth rates upon $\tau$ for varying inlet temperatures. The solid symbols stand for the operating conditions, while the hollow symbols are hypothetical conditions to show a wider range of the trends.}
\label{oscilos} 
\end{figure}

\subsection{The impact of the spray}
The impact of the spray dynamics and droplets evaporation on the observed bifurcation behaviour is now discussed. 

In the pilot stage, the spray characteristics are mainly affected by the fuel supply pressure which is kept almost constant. Furthermore, the fuel droplets enter into the combustion chamber immediately after being ejected. For these reasons it is possible to conclude that the impact of the inlet temperature on the pilot stage is negligible. The whole process of spray in the main stage can be simplified in two parts: a) the injection of fuel, b) the convection and vaporisation of the fuel. Based on the summaries of Mellor~\cite{mellor1990design}, an emprical equation is used to estimate the initial Sauter mean diameter (SMD) of droplet $d_{ini}$:
\begin{equation}
d_{ini}=0.73(  \rho_{a}d_{inj})^{0.5} ( \frac{ \sigma_{l}}{\rho_{a}U^2_{\theta}})^{0.5}+0.04 d_{inj}^{0.55} ( \frac{ \mu^2_{l}}{ \sigma _{l}\rho_{l}} )^{0.45}
\label{SMD}
\end{equation}
where subscripts $a$ stands for inlet air while $l$ stands for liquid fuel. In Eq.~\ref{SMD}, $d_{inj}$ is the diameter of the fuel injector (0.5 mm), $\rho_{a}$ and $U_{\theta}$ are the density and velocity magnitude of the inlet air. Please note due to the swirl effect, the velocity magnitude that the fuel column faces is determined by $U_{\theta}=U_{m}/\cos\theta$ , where $U_{m}$ is the axial velocity inside the main stage and $\theta$ is the angle of the swirler vanes ($60^\circ$). The other variables $\sigma_{l}$, $\rho_l$ and $\mu_{l}$ are the surface tension (0.027 N/m), density (780 kg/m$^3$) and dynamic viscosity (0.001 Pa/s) of the liquid fuel, respectively.

It is then necessary to estimate the evaporation time $\tau_e$ of the liquid droplets. A classical $d^2$ law with convection correlation is used~\cite{ranz1952evaporation} here as 

\begin{equation}
d^2_{ini}-d^2_{exit}=\tau_m Nu^*K_e
\label{tau}
\end{equation}
where $d^2_{exit}$ is the fuel droplet diameter at the exit of the main stage. $K_e$ is the evaporation coefficient while $Nu^*$ is Nusselt number, which can be estimated by the well-known Frossling correlations~\cite{froessling1968evaporation}. The equations for the above two parameters are as follows:
\begin{equation}
K_e= \frac{8 \lambda_a}{\rho_ac_p}ln(1+B) 
\label{Ke}
\end{equation}

\begin{equation}
Nu^*=1+0.276Re^{0.5}Pr^{0.33}
\label{Nu}
\end{equation}
where $\lambda_a$ and $c_p$ is the thermal conductivity and specific heat capacity of air, respectively. The Spalding transfer number $B=c_p(T_a-T_l)/L_v$ is expressed regarding the temperature of air $T_a$ and liquid fuel $T_l$ and latent heat $L_v$ (291 kJ/kg). It is standard to evaluate the air properties by the ``1/3 rule'' with the reference temperature given by $\bar{T}=1/3T_a+2/3T_l$. Here the liquid fuel temperature $T_l$ is fixed at $300$ K. Additive convection effects are taken into consideration by Eq.~\ref{Nu}, where the Reynolds number $Re=\rho_a U_m d_l/\mu_a$ and Prandtl number $Pr=\mu_a c_p/\lambda_a$.

Finally, the evaporation ratio $R_{evap}$ can be calculated as:
\begin{equation}
R_{evap}=\frac{\dot{m}_e}{\dot{m}_f}=1-\frac{d^3_{exit}}{d^3_l}
\label{evaporation}
\end{equation}

\begin{figure}[htbp]
\centering 
\includegraphics[width=0.5\textwidth]{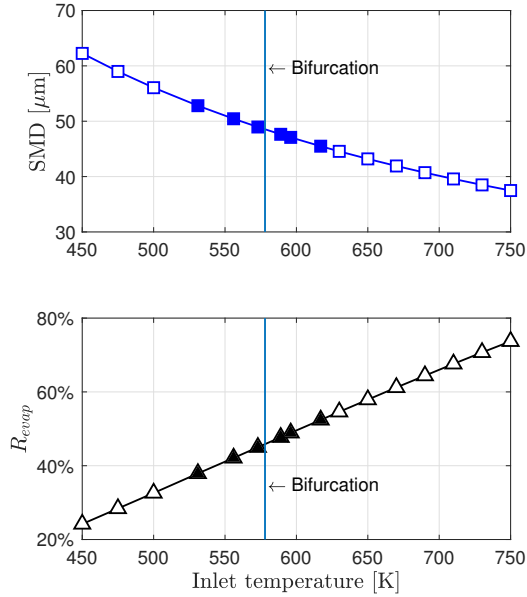}
\caption{Results of estimated spray properties with different inlet temperature. The square and triangle symbols represent the SMD of the initial droplet and the evaporation ratio $R_{evap}$. The solid symbols stand for the operating conditions, while the hollow symbols are hypothetical conditions to show a wider range of the trends.}
\label{droplet}
\end{figure}

The estimated spray properties with different inlet temperatures ranging from 450 K to 750 K are shown in Fig.~\ref{droplet}. The square and triangle symbols represent the SMD of the initial droplet and the evaporation ratio $R_{evap}$. The solid symbols stand for the operating conditions, while the hollow symbols are hypothetical conditions to show a wider range of the trends. With the given geometry, the inlet temperature has significant effects on the spray properties. With the increase of inlet temperature $T_a$, the initial SMD of the fuel droplet decreases, while the evaporation ratio $R_{evap}$ increases linearly. This means that with a higher $T_a$, more fuel is evaporated at the exit of swirler enhancing the premixing process happening in this zone of the system. The consequential reduction of the oscillations of the equivalence ratio, one of the instability driving mechanisms can then also be assumed to be one of the reasons leading to the system stabilisation.

%%%%%%%%%%%%%%%%%%%%%
%                                 %
%.           CONCLUSIONS                 %
%                                 %
%%%%%%%%%%%%%%%%%%%%%

\section{Conclusions}
\vspace*{0mm}

The present article reports experimental observation and analysis of a supercritical bifurcation of combustion instabilities triggered by inlet temperature. The studies are performed with a kerosene-fuelled LPP combustor under elevated temperature and pressure conditions. The inlet temperature varies from 530 K to 617 K while all the other parameters keep fixed. The combustor features high amplitude oscillations at the low-temperatures from 530 K to 556 K but becomes thermoacoustically stable suddenly once the inlet temperature exceeds 570 K and remains stable beyond that. Repeated experiments further confirm that this temperature triggered-bifurcation is supercritical without hysteresis behaviour.

The oscillation characteristics are carefully demonstrated with signals sampled from different sensors, as well as the flame dynamics captured by a high-speed camera with a CH* filter. For the unstable cases, the flames feature a periodic axial motion of lift-off and re-ignition, which is also known as Helmholtz mode or bulk mode. The analysis shows that the phase difference between chemiluminescence and pressure signals increases with the inlet temperature, and exceeds $90^\circ$ when the temperature is higher than 570 K. The phase correlation between the heat release rate and pressure is therefore thought to be the major reason for the observed bifurcation. 

A thermoacoustic analysis is conducted with a low order network model to illustrate the effects of convection time on growth rate and frequency. With the increase of temperature, the estimated convection time decreases, resulting in a higher frequency and lower growth rate (indicating system to be stable). This agrees with the experimental observation. The effects of inlet temperature on spray properties are further examined by the estimation of SMD and the evaporation ratio of fuel droplets. With a higher $T_a$, more fuel is evaporated at the exit of swirler, enhancing the premixing process happening in this zone of the system. The consequential reduction of the oscillations of the equivalence ratio can then also be assumed to be one of the reasons leading to the system stabilisation.

This study emphasizes the extreme sensitivity of the combustion instabilities to the inlet temperature and convection time in a complex spray LPP combustor. It could also help to improve the design of LPP combustors.

\section{Acknowledgment}
This work was financially supported by the National Natural Science Foundation of China (91641109, 51606004) and the European Research Council (grant no.772080) via the ERC Consolidator Grant AFIRMATIVE (2018-23).

%\begin{thebibliography}{00}


\bibliographystyle{elsarticle-num}

\bibliography{References.bib}

\begin{thebibliography}{10}
\expandafter\ifx\csname url\endcsname\relax
  \def\url#1{\texttt{#1}}\fi
\expandafter\ifx\csname urlprefix\endcsname\relax\def\urlprefix{URL }\fi
\expandafter\ifx\csname href\endcsname\relax
  \def\href#1#2{#2} \def\path#1{#1}\fi

\bibitem{li2016emission}
L.~Li, Y.~Lin, Z.~Fu, C.~Zhang, Emission characteristics of a model combustor
  for aero gas turbine application, Exp. Therm. Fluid Sci. 72 (2016) 235--248.

\bibitem{rayleigh1878explanation}
L.~Rayleigh, The explanation of certain acoustical phenomena, Roy. Inst. Proc.
  8 (1878) 536--542.

\bibitem{lieuwen2005combustion}
T.~C. Lieuwen, V.~Yang, Combustion instabilities in gas turbine engines:
  operational experience, fundamental mechanisms, and modeling, American
  Institute of Aeronautics and Astronautics, 2005.

\bibitem{al2015review}
Y.~M. Al-Abdeli, A.~R. Masri, Review of laboratory swirl burners and
  experiments for model validation, Exp. Therm. Fluid Sci. 69 (2015) 178--196.

\bibitem{ebi2017flame}
D.~{Ebi}, A.~{Denisov}, G.~{Bonciolini}, E.~{Boujo}, N.~{Noiray}, Flame
  dynamics intermittency in the bi-stable region near a subcritical hopf
  bifurcation, J. Eng. Gas Turbines Power 140 (2017) 61504.

\bibitem{han2019flame}
X.~Han, D.~Laera, A.~S.~Morgans, C.-J. Sung, X.~Hui, Y.~Lin, Flame
  macrostructures and thermoacoustic instabilities in stratified swirling
  flames, Proc. Combust. Inst. 37 (2019) 5377--5384.

\bibitem{yoon2013effect}
J.~Yoon, M.-K. Kim, J.~Hwang, J.~Lee, Y.~Yoon, Effect of fuel--air mixture
  velocity on combustion instability of a model gas turbine combustor, Appl.
  Therm. Eng. 54 (2013) 92--101.

\bibitem{kim2019experimental}
J.~Kim, M.~Jang, K.~Lee, A.~Masri, Experimental study of the beating behavior
  of thermoacoustic self-excited instabilities in dual swirl combustors, Exp.
  Therm. Fluid Sci. (2019).

\bibitem{broda1998experimental}
J.~Broda, S.~Seo, R.~Santoro, G.~Shirhattikar, V.~Yang, An experimental study
  of combustion dynamics of a premixed swirl injector, Proc. Combust. Inst. 27
  (1998) 1849--1856.

\bibitem{huang2004bifurcation}
Y.~Huang, V.~Yang, Bifurcation of flame structure in a lean-premixed
  swirl-stabilized combustor: transition from stable to unstable flame,
  Combust. Flame 136 (2004) 383--389.

\bibitem{huang2009dynamics}
Y.~Huang, V.~Yang, Dynamics and stability of lean-premixed swirl-stabilized
  combustion, Prog. Energy Combust. Sci. 35 (2009) 293--364.

\bibitem{strogatz2018nonlinear}
S.~H. Strogatz, Nonlinear dynamics and chaos: with applications to physics,
  biology, chemistry, and engineering, CRC Press, 2018.

\bibitem{Kashinath2014Nonlinear}
K.~Kashinath, I.~C. Waugh, M.~P. Juniper, Nonlinear self-excited thermoacoustic
  oscillations of a ducted premixed flame: Bifurcations and routes to chaos, J.
  Fluid Mech. 761 (2014) 399--430.

\bibitem{kabiraj2012bifurcations}
L.~Kabiraj, R.~I. Sujith, P.~Wahi, Bifurcations of self-excited ducted laminar
  premixed flames, J. Eng. Gas Turbines Power 134 (2012) 031502.

\bibitem{weng2016investigation}
F.~Weng, S.~Li, D.~Zhong, M.~Zhu, Investigation of self-sustained beating
  oscillations in a {R}ijke burner, Combust. Flame 166 (2016) 181--191.

\bibitem{prieur2017hysteresis}
K.~Prieur, D.~Durox, T.~Schuller, S.~Candel, A hysteresis phenomenon leading to
  spinning or standing azimuthal instabilities in an annular combustor,
  Combust. Flame 175 (2017) 283--291.

\bibitem{Moeck2008Subcritical}
J.~Moeck, M.~Bothien, S.~Schimek, A.~Lacarelle, C.~Paschereit, Subcritical
  thermoacoustic instabilites in a premixed combustor, in: 14th AIAA/CEAS, AIAA
  2008-2946, 2008.

\bibitem{laera2017finite}
D.~Laera, G.~Campa, S.~Camporeale, A finite element method for a weakly
  nonlinear dynamic analysis and bifurcation tracking of thermo-acoustic
  instability in longitudinal and annular combustors, Appl. Energy 187 (2017)
  216--227.

\bibitem{janus1996model}
M.~C. Janus, G.~A. Richards, A model for premixed combustion oscillations,
  Tech. rep., USDOE Morgantown Energy Technology Center, WV (United States)
  (1996).

\bibitem{Noiray2008A}
N.~Noiray, D.~Durox, T.~Schuller, S.~Candel, A unified framework for nonlinear
  combustion instability analysis based on the describing function, J. Fluid
  Mech. 615 (2008) 139--167.

\bibitem{lieuwen2001mechanism}
T.~Lieuwen, H.~Torres, C.~Johnson, B.~Zinn, A mechanism of combustion
  instability in lean premixed gas turbine combustors, J. Eng. Gas Turbines
  Power 123 (2001) 182--189.

\bibitem{giuliani2002influence}
F.~Giuliani, P.~Gajan, O.~Diers, M.~Ledoux, Influence of pulsed entries on a
  spray generated by an air-blast injection device: An experimental analysis on
  combustion instability processes in aeroengines, Proc. Combust. Inst. 29
  (2002) 91--98.

\bibitem{de2009investigations}
M.~de~la Cruz~Garc{\'\i}a, E.~Mastorakos, A.~Dowling, Investigations on the
  self-excited oscillations in a kerosene spray flame, Combust. Flame 156
  (2009) 374--384.

\bibitem{tachibana2015experimental}
S.~Tachibana, K.~Saito, T.~Yamamoto, M.~Makida, T.~Kitano, R.~Kurose,
  Experimental and numerical investigation of thermo-acoustic instability in a
  liquid-fuel aero-engine combustor at elevated pressure: Validity of
  large-eddy simulation of spray combustion, Combust. Flame 162 (2015)
  2621--2637.

\bibitem{sidey2018stabilisation}
J.~A. Sidey, E.~Mastorakos, Stabilisation of swirling dual-fuel flames, Exp.
  Therm. Fluid Sci. 95 (2018) 65--72.

\bibitem{providakis2012characterization}
T.~Providakis, L.~Zimmer, P.~Scouflaire, S.~Ducruix, Characterization of the
  acoustic interactions in a two-stage multi-injection combustor fed with
  liquid fuel, J. Eng. Gas Turbines Power 134 (2012) 111503.

\bibitem{renaud2015flame}
A.~Renaud, S.~Ducruix, P.~Scouflaire, L.~Zimmer, Flame shape transition in a
  swirl stabilised liquid fueled burner, Proc. Combust. Inst 35 (2015)
  3365--3372.

\bibitem{renaud2017bistable}
A.~Renaud, S.~Ducruix, L.~Zimmer, Bistable behaviour and thermo-acoustic
  instability triggering in a gas turbine model combustor, Proc. Combust. Inst
  36 (2017) 3899--3906.

\bibitem{temme2014combustion}
J.~E. Temme, P.~M. Allison, J.~F. Driscoll, Combustion instability of a lean
  premixed prevaporized gas turbine combustor studied using phase-averaged
  {PIV}, Combust. Flame 161 (2014) 958--970.

\bibitem{dhanuka2011lean}
S.~K. Dhanuka, J.~E. Temme, J.~F. Driscoll, Lean-limit combustion instabilities
  of a lean premixed prevaporized gas turbine combustor, Proc. Combust. Inst.
  33 (2011) 2961--2966.

\bibitem{han2017combustion}
X.~Han, X.~Hui, C.~Zhang, Y.~Lin, P.~He, C.-J. Sung, Combustion instabilities
  in a lean premixed pre-vaporized combustor at high-pressure high-temperature,
  in: ASME Turbo Expo 2017, GT2017-65190, 2017.

\bibitem{mao2019experimental}
Y.~Mao, L.~Yu, Z.~Wu, W.~Tao, S.~Wang, C.~Ruan, L.~Zhu, X.~Lu, Experimental and
  kinetic modeling study of ignition characteristics of rp-3 kerosene over
  low-to-high temperature ranges in a heated rapid compression machine and a
  heated shock tube, Combust. Flame 203 (2019) 157--169.

\bibitem{han2016effect}
X.~Han, X.~Hui, H.~Qin, Y.~Lin, M.~Zhang, C.-J. Sung, Effect of the diffuser on
  the inlet acoustic boundary in combustion-acoustic coupled oscillation, in:
  ASME Turbo Expo 2016, GT2016-57046, 2016.

\bibitem{hardalupas2004local}
Y.~Hardalupas, M.~Orain, Local measurements of the time-dependent heat release
  rate and equivalence ratio using chemiluminescent emission from a flame,
  Combust. Flame 139 (2004) 188--207.

\bibitem{tsuboi1985light}
T.~Tsuboi, K.~Inomata, Y.~Tsunoda, A.~Isobe, K.-i. Nagaya, Light absorption by
  hydrocarbon molecules at 3.392 $\mu$m of {H}e-{N}e laser, Jpn. J. Appl. Phys.
  24 (1985) 8.

\bibitem{abarbanel1993analysis}
H.~D. Abarbanel, R.~Brown, J.~J. Sidorowich, L.~S. Tsimring, The analysis of
  observed chaotic data in physical systems, Rev. Mod. Phys. 65 (1993) 1331.

\bibitem{guan2018nonlinear}
Y.~Guan, P.~Liu, B.~Jin, V.~Gupta, L.~K. Li, Nonlinear time-series analysis of
  thermoacoustic oscillations in a solid rocket motor, Exp. Therm. Fluid Sci.
  98 (2018) 217--226.

\bibitem{ahn2018low}
B.~Ahn, J.~Lee, S.~Jung, K.~T. Kim, Low-frequency combustion instabilities of
  an airblast swirl injector in a liquid-fuel combustor, Combust. Flame 196
  (2018) 424--438.

\bibitem{dowling2003acoustic}
A.~P. Dowling, S.~R. Stow, Acoustic analysis of gas turbine combustors, J.
  Propul. Power 19 (2003) 751--764.

\bibitem{schuller2003unified}
T.~Schuller, D.~Durox, S.~Candel, A unified model for the prediction of laminar
  flame transfer functions: comparisons between conical and v-flame dynamics,
  Combust. Flame 134 (2003) 21--34.

\bibitem{crocco1956theory}
L.~Crocco, S.-I. Cheng, Theory of combustion instability in liquid propellant
  rocket motors, Tech. rep., Princeton Univ. NJ (1956).

\bibitem{hermeth2014bistable}
S.~Hermeth, G.~Staffelbach, L.~Y. Gicquel, V.~Anisimov, C.~Cirigliano,
  T.~Poinsot, Bistable swirled flames and influence on flame transfer
  functions, Combust. Flame 161 (2014) 184--196.

\bibitem{lieuwen1998role}
T.~Lieuwen, B.~T. Zinn, The role of equivalence ratio oscillations in driving
  combustion instabilities in low {N}{O}x gas turbines, Proc. Combust. Inst. 27
  (1998) 1809--1816.

\bibitem{poinsot2005theoretical}
T.~Poinsot, D.~Veynante, Theoretical and numerical combustion, RT Edwards,
  Inc., 2005.

\bibitem{li2015time}
J.~Li, A.~S. Morgans, Time domain simulations of nonlinear thermoacoustic
  behaviour in a simple combustor using a wave-based approach, J. Sound Vib.
  346 (2015) 345--360.

\bibitem{yang2019systematic}
D.~Yang, D.~Laera, A.~S. Morgans, A systematic study of nonlinear coupling of
  thermoacoustic modes in annular combustorscombutors, J. Sound Vib. 456 (2019)
  137--161.

\bibitem{han2015openfoam}
X.~Han, J.~Li, A.~S. Morgans, Prediction of combustion instability limit cycle
  oscillations by combining flame describing function simulations with a
  thermoacoustic network model, Combust. Flame 162 (2015) 3632--3647.

\bibitem{li2017numerical}
J.~Li, Y.~Xia, A.~S. Morgans, X.~Han, Numerical prediction of combustion
  instability limit cycle oscillations for a combustor with a long flame,
  Combust. Flame 185 (2017) 28--43.

\bibitem{xia2019numerical}
Y.~Xia, D.~Laera, W.~P. Jones, A.~S. Morgans, Numerical prediction of the
  {F}lame {D}escribing {F}unction and thermoacoustic limit cycle for a
  pressurized gas turbine combustor, Combust. Sci. Technol. (2019) 979--1002.

\bibitem{davies1988practical}
P.~Davies, Practical flow duct acoustics, J. Sound Vib. 124 (1988) 91--115.

\bibitem{marble1977acoustic}
F.~Marble, S.~Candel, Acoustic disturbance from gas non-uniformities convected
  through a nozzle, J. Sound Vib. 55 (1977) 225--243.

\bibitem{dowling1997nonlinear}
A.~P. Dowling, Nonlinear self-excited oscillations of a ducted flame, J. Fluid
  Mech. 346 (1997) 271--290.

\bibitem{laera2017weakly}
D.~Laera, S.~M. Camporeale, A weakly nonlinear approach based on a distributed
  flame describing function to study the combustion dynamics of a full-scale
  lean-premixed swirled burner, J. Eng. Gas Turb. Power 139 (2017) 091501.

\bibitem{mellor1990design}
A.~Mellor, Design of modern turbine combustors, Academic Pr., 1990.

\bibitem{ranz1952evaporation}
W.~Ranz, W.~R. Marshall, et~al., Evaporation from drops, Chem. Eng. Prog. 48
  (1952) 141--146.

\bibitem{froessling1968evaporation}
N.~Froessling, On the evaporation of falling drops, Tech. rep., Army Biological
  Labs Frederick MD (1968).

\end{thebibliography}
%\bibliography{/Users/hanxiao/Onedrive/A_Submission/XiaoHan}

%\bibitem{}

%\end{thebibliography}
\end{document}